\documentclass{aa}  
\usepackage{graphicx}
\usepackage{txfonts}
\usepackage{newtxtext,newtxmath}
\usepackage[T1]{fontenc}
\usepackage{ae,aecompl}
\usepackage{graphicx}   
\usepackage{amsmath}    
\usepackage{hyperref}
\usepackage{multirow}
\usepackage{threeparttable}
\usepackage[export]{adjustbox}
\usepackage{fontawesome}\usepackage[dvipsnames]{xcolor}
\usepackage{txfonts}
\definecolor{BlueGreen}{rgb}{0.0, 0.58, 0.71}
\begin{document}

   \title{Physical properties of circumnuclear ionising clusters\\ III. Kinematics of gas and stars in NGC\,7742}

   \author{S. Zamora
          \inst{1} \inst{2}\fnmsep\thanks{PhD fellow of Ministerio de Educación y Ciencia, Spain, BES-2017-080509, CEAL-AL/2017-02}
          \and
          A. I. Díaz \inst{1}\inst{2}
          \and
          Roberto Terlevich \inst{3}\inst{4}\inst{5}
          \and
          Elena Terlevich \inst{3} \inst{5}
          \and
          R. Amorín \inst{6}\inst{7}
          }

   \institute{Departamento de Física Teórica, Universidad Autónoma de Madrid, 28049 Madrid, Spain
         \and
             CIAFF, Universidad Autónoma de Madrid, 28049 Madrid, Spain
        \and 
            Instituto de Astrofísica, Óptica y Electrónica, 72840  Puebla, México
        \and
            Institute of Astronomy, University of Cambridge, Madingley Road, Cambridge, UK
        \and
        Facultad de Astronomia y Geofisica, Universidad de La Plata, La Plata, Argentina
        \and
            Instituto de Investigación Multidisciplinar en Ciencia y Tecnología, Universidad de La Serena, La Serena, Chile.
        \and
             Departamento de Astronomía, Universidad de La Serena, La Serena, Chile.
             }

   \date{ }

 
  \abstract
   {}
   {
   For this third paper in the series, we studied the kinematics of the ionised gas and stars, and calculated the dynamical masses of the circumnuclear star-forming regions in the ring  of the face-on spiral NGC~7742.  
   }
  {
  We  used high spectral resolution data from the MEGARA instrument mounted on the Gran Telescopio Canarias (GTC) to measure the kinematical components of the nebular emission lines of selected HII regions and the stellar velocity dispersions from the CaT absorption lines that allow the derivation of the associated cluster virialised masses. 
   }
   {
    The emission line profiles show two different kinematical components: a narrow one with a velocity dispersion of $\sim$ 10 km/s and a broad one with a velocity dispersion similar to the values found for the stellar absorption lines. The derived star cluster dynamical masses range from 2.5 $\times$ 10$^6$ to 10.0 $\times$ 10$^7$ M$_\odot$.
   }
   {
   The comparison of gas and stellar velocity dispersions suggests a scenario where the clusters have formed simultaneously in a first star formation episode with a fraction of the stellar evolution feedback remaining trapped in the cluster, subject to the same gravitational potential as the cluster stars. Between 0.15 and 7.07 \% of the total dynamical mass of the cluster would have cooled down and formed a new, younger, population of stars, responsible for the ionisation of the gas currently observed.
   }

   \keywords{ISM: abundances -- ISM: H II regions -- galaxies: ISM -- galaxies: starburst -- galaxies: star clusters -- galaxies: star formation}

   \maketitle
%

\section{Introduction}
Observations of central regions have revealed that nuclear rings are present in at least 20 \% of spiral galaxies, with this percentage increasing in galaxies hosting an active galactic nucleus (AGN) \citep[see][]{2005A&A...429..141K}. These structures, typically with diameters of hundreds of parsecs, exhibit high star formation rates that have been studied by several authors over the past decades across different wavelengths \citep[e.g.][]{2022ApJ...935...19L, 2023ApJ...942..108L, 2012A&A...543A..61B, 2008AJ....135..479B, 2008ApJS..174..337M}. These star-forming regions, often referred to as hotspots or circumnuclear star-forming regions (CNSFRs), are similar to luminous and large disk HII regions, but look more compact and show higher peak surface brightness \citep{1989AJ.....97.1022K}. In early-type  spirals they are expected to be of higher metallicity as it corresponds to their position near the galactic bulge \citep{2007MNRAS.382..251D}. In many cases, CNSFRs contribute substantially to the emission of the entire nuclear region seen in the presence of an active nucleus. Their high H$\alpha$ luminosities, typically higher than 10$^{39}$ erg s$^{-1}$, overlap with those of HII galaxies, and  point to relatively massive star clusters as their ionisation source. Hence, these regions are excellent places to study how star formation proceeds in circumnuclear environments; they are crucial for understanding the connection between nuclear activity and the star formation processes occurring in these hostile environments \citep[see][]{2007MNRAS.380..949S}.

In previous works, \citet{2007MNRAS.378..163H,2009MNRAS.396.2295H,2010MNRAS.402.1005H} analysed long-slit high spectral resolution data of  CNSFRs in three nearby galaxies: NGC~3351, NGC~2903, and NGC~3310. They used the 4.2m \textit{William Herschel} Telescope (WHT) and the ISIS spectrograph in the blue and red configurations, which provides a comparable velocity resolution of about 13 km/s in both spectral ranges. The H$\beta$ and [OIII]$\lambda$5007 \AA\ emission lines and the CaII triplet lines (CaT) were present in the spectra, thus allowing the measurement of the gas and star velocity dispersions,  which, together with their sizes measured on the WPC2-HST camera and assuming virialisation, made possible the derivation of the dynamical masses of the young clusters in the regions. These masses, found to be between 4.9 $\times $ 10$^6$ and 1.9 $\times$ 10$^8$ M$_\odot $, seem to be rather high and  suggest star-forming complexes rather than individual clusters. However, the masses of the corresponding ionising clusters, as derived from H$\alpha$ luminosities under the assumption of no photon escape, only accounts for between 1 \% and 10 \% of the derived dynamical mass, which raises the question of the possible presence of more than one stellar population in these massive circumnuclear star clusters. 

On the other hand, one of the main results of these works shows that for the three observed galaxies, the gas velocity dispersions as measured from hydrogen recombination lines (H$\beta$ and Paschen emission lines) are found to be considerably smaller (by about 25 km/s) than those measured from the stellar CaT lines. In turn, the velocity dispersion obtained from the [OIII]$\lambda $5007 \AA\ line is very close to the measured stellar value. This result suggests the possible existence of two kinematically distinct components in the gas in all regions, although not all of them are associated with the ionising clusters \citep[see][]{2013MNRAS.432..810H}. 

The best possibility to resolve this question is to gather observations with both high spatial and spectral resolution simultaneously. At present, the ideal instrument for this is MEGARA, a high-resolution integral field spectrograph (IFS) installed on the 10.4m GTC telescope, covering a field of view (FoV) of 12.5 × 11.3 arcsec$^2$ with a spatial resolution of 0.62 arcsec and a spectral range from 3650 to 9700 \AA\ at three different resolutions, R $ \sim$ 6000, 12000, and 20000. The MEGARA spatial and spectral resolution characteristics ensure the proper identification and resolution of the ionising clusters present in CNSFRs so that the derived masses obtained can be meaningful and reliable.

In the first paper of this series we   analyse CNSFRs in the galaxy NGC~7742 using IFS data obtained with the MUSE spectrograph, which provides spectral resolutions between  R $\sim$ 1770 - 3590 at 4800 and 9300 \AA,\ respectively. 
This galaxy is classified as an SA(r)b galaxy and its morphology is dominated by a circumnuclear ring located at around 1 kpc from the galaxy centre. The galaxy harbours a mildly active nucleus whose spectrum is dominated by emission lines characteristic of both LINER and HII regions. In total, 88 ring HII regions with H$\alpha$ fluxes larger than 10$^{-18}$erg/s/cm$^2$ were selected and characterised using their nebular spectra with the help of photoionisation and stellar population synthesis models (Cloudy and PopStar, respectively). These regions seem to be ionised by young clusters with mean ages of about 5 Ma. The chemical composition of the ionised gas was calculated using sulphur as a tracer and were found to be between 0.25 and 2.4 times the solar value, with most regions showing values slightly below solar. This result is consistent with an effective temperature of the cluster stars between 35000 K and 40000 K, as estimated from the ratio of HeI to H ionising photons. 
The high spatial resolution offered by MUSE allowed the measurement of the CNSFR sizes on the H$\alpha$ flux maps, all of which were spatially resolved. These measured sizes are fully consistent with those predicted by photoionisation models, thus leaving little room for photon escape and justifying the assumption of the regions being radiation bounded. The young ionising star clusters powering this radiation have masses of around 3.5 $\times$ 10$^4$ M$_{\odot}$, comparable to the mass of ionised gas and about 19 \% of the corrected photometric cluster mass that includes a young  non-ionising stellar population whose estimated age is around 300 Ma. A minor merger scenario is proposed for the formation of these clusters due to the counter rotating nature of the ring \citep{2004A&A...424..447P,2002MNRAS.329..513D} and the homogeneity of abundances and continuum colours.


The present work is focused on three principal objectives: first, testing the capabilities of the new MEGARA instrument for the study of resolved kinematical components in the ionised gas of HII regions; second, determining the kinematic origin of such components if present, for which observing both recombination lines (H$\alpha$) and collisionally excited lines ([NII] and [SIII] is needed; and third, measuring the stellar velocity dispersion along the line of sight from the CaT lines in order to derive cluster dynamical masses. Accomplishing these objectives will ensure the cluster origin of the CaT lines and the reliability of this derivation. It should be said that a proper comparison between ionising and dynamical masses for ionising young star clusters has not yet been done and is much needed.

In Sect. 2 we describe the observations and the reduction of the data, focusing on the new MEGARA observations. Section 3 is devoted to the presentation of our results. The discussion of cluster sizes and dynamical masses and the comparison between gas and star kinematical components is presented in Sect. 4. Finally, in Sect. 5 we summarise our work and present our conclusions.

\section{Observations and data reduction}\label{sec:observations:MEGARA}
For this work we   used data for the ring HII regions of \href{https://ned.ipac.caltech.edu/byname?objname=NGC7742&hconst=67.8&omegam=0.308&omegav=0.692&wmap=4&corr_z=1}{NGC 7742} \citep[J2000, RA=356.065542 deg, DEC=10.767083 deg,][]{2006AJ....131.1163S} obtained with the high-resolution integral field spectrograph MEGARA \citep[Multi-Espectrógrafo en GTC de Alta Resolución para Astronomía,][]{2018SPIE10702E..16C} attached to the GTC 10.4 m telescope at La Palma Observatory. We   used MEGARA in  the integral-field Unit (IFU) mode working in the visible wavelength range from 3655 to 9700 \AA\ and providing a high spectral resolution (up to R$_{FWHM}$ = $\lambda / \Delta \lambda$ $\sim$  20000) and spatial resolution (0.62 arcsec spaxel size) simultaneously in a moderate size field of view (FoV) (12.5 $\times$ 11.3 arcsec$^2$). It has a total of 623 hexagonal fibres  distributed in 567 science-fibres and 56 sky-fibres located at the edges of the FoV at distances from 1.7 arcmin to 2.5 arcmin from the centre of the fibre bundle.


\begin{figure}
\includegraphics[width=\columnwidth]{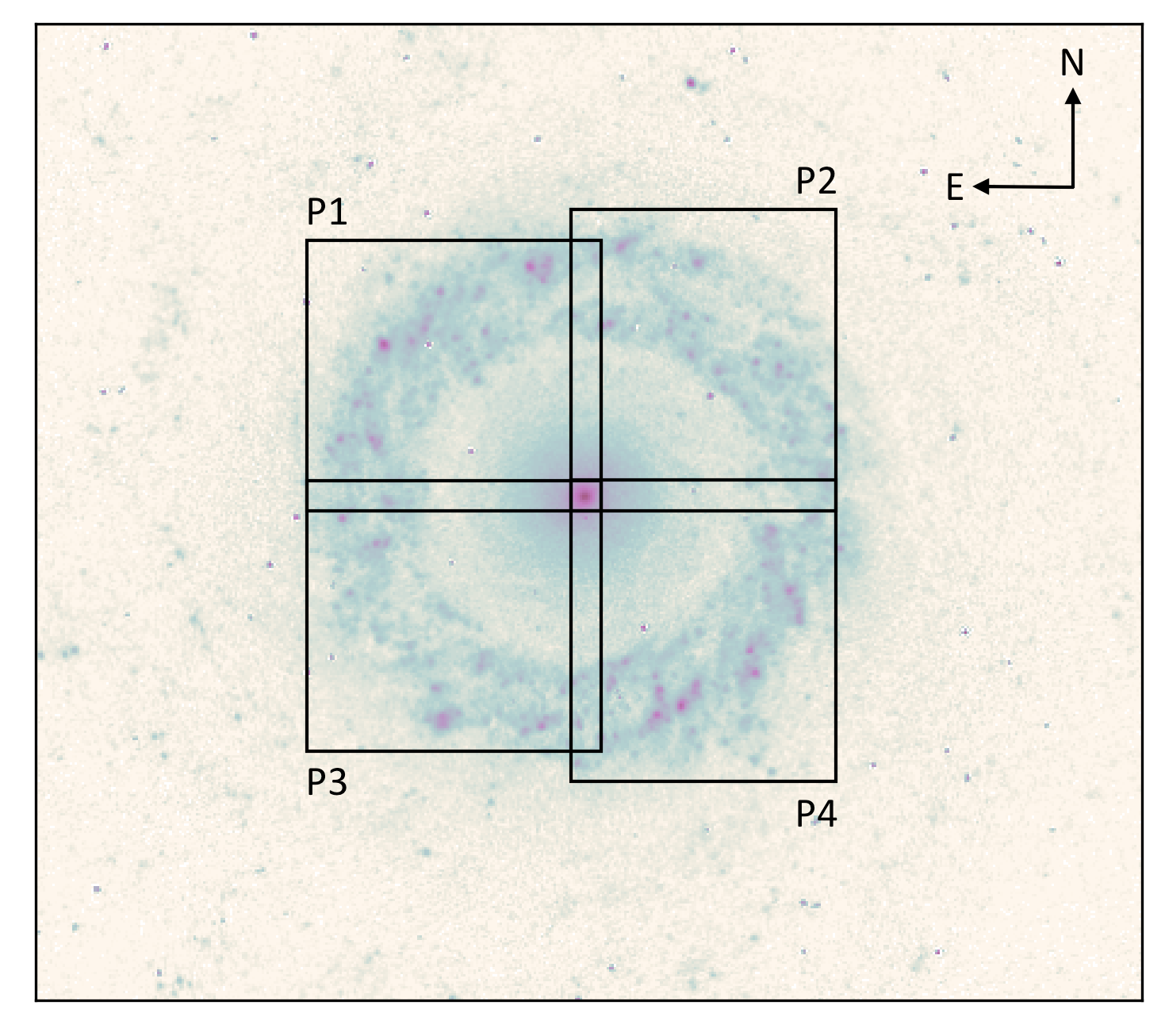}
\caption{WPC2-HST image in the F336W filter showing the circumnuclear region of NGC7742 and the MEGARA fields of view. North is at the top and east is to the left.}
\label{fig:observations}
\end{figure}

The observations were acquired during three nights on 2021 August 28 and  31, and September 14\footnote{Proposal GTC28-21A: \textit{A high spatial and spectral resolution study of circumnuclear star-forming regions}\\ PI: Ángeles I. Díaz. Semester 2021-A.} under spectroscopic conditions. The seeing during the observations ranged between approximately 0.7 and 0.9 arcsec and the airmass of the science images was always less than 1.25. Four pointings were needed to cover the whole circumnuclear ring extension of NGC 7742 (approximately the central 24x24 arcsec$^2$; see Fig. \ref{fig:observations}).

\begin{table*}
\centering
\caption{Target observations.}
\label{tab:targets}
\begin{tabular}{ccccccc}
\hline
Target      & \begin{tabular}[c]{@{}c@{}}RA \\ (hh:mm:ss.ss)\end{tabular}& \begin{tabular}[c]{@{}c@{}}DEC \\ (dd:mm:ss.ss)\end{tabular} &\begin{tabular}[c]{@{}c@{}}Field of View \\ Position Angle$^a$ \\ (degrees)\end{tabular} & \begin{tabular}[c]{@{}c@{}}Exp. Time$^b$ (s)\\ VPH863\_HR \end{tabular}  & \begin{tabular}[c]{@{}c@{}}Exp. Time$^b$ (s)\\ VPH890\_LR\end{tabular} & \begin{tabular}[c]{@{}c@{}}Exp. Time$^b$ (s)\\ VPH665\_HR\end{tabular}\\ \hline 
NGC7742\_P1 & 23:44:16.02 &     10:46:05.80 & 0 & 1050 & 450 & 684\\
NGC7742\_P2 & 23:44:15.28 & 10:46:06.40 & 0 & 1050 & - & 684\\
NGC7742\_P2 & 23:44:15.28 & 10:46:06.40 & 90 & 1050 & 450 & 342\\
NGC7742\_P3 & 23:44:16.02 & 10:45:55.60 & 0 & 1050 & 450 & 684\\
NGC7742\_P4 & 23:44:15.28 & 10:45:55.00 & 90 & 1050 & 450 & 342\\ \hline
\end{tabular}
\tablefoot{
$^a$ From north towards east. The orientation of 0 deg implies that the north is aligned with the short side and the east--west axis with the long side of the detector.
$^b$ Total target observed time without overheads.
}
\end{table*}

Three dispersion elements (VPHs) were used: VPH863\_HR (HR-I, R$_{FWHM}$ $\sim$ 20500), including the calcium triplet stellar lines (CaT) at high spectral resolution;  VPH890\_LR (LR-Z, R$_{FWHM}$ $\sim$ 5800), including the  [SIII]$\lambda \lambda$9069,9532 \AA\ emission lines and the CaT lines at low resolution; and VPH665\_HR (HR-R, R$_{FWHM}$ $\sim$ 20050), including the [NII]$\lambda \lambda$6548,84 \AA,  H$\alpha$, and [SII]$\lambda \lambda$6717,32 \AA\ emission lines at high resolution. A relation of the total integration times and their splitting in different exposures for each of the used  set-ups is given in Table \ref{tab:targets}. 
Using these configurations, velocity dispersions of around 6.4 km/s and 22.5 km/s in full width at half maximum (FWHM) were expected at high and low spectral resolutions, respectively.

Calibration images were acquired on the observing nights and on the preceding and following nights. They consisted of bias frames to create a CCD spatial response function;  flat-field frames necessary for locating each fibre spectrum on the CCD and for performing a spatial instrumental response function; arc calibration exposures needed for wavelength calibration; and three spectrophotometric standard star exposures needed for performing a spectral instrumental response function. The standard procedures of bias subtraction, tracing and extraction of each fibre spectra, flat-field correction, and wavelength calibration were followed applying some required modifications to the standard MEGARA pipeline  \citep{2018zndo...2206856P}\footnote{see \url{https://github.com/guaix-ucm/megaradrp/}} using the calibration files present in \citet{2018zndo...2270518C}. Flux calibration was not needed for our work, and we decided not to do it in order not to introduce additional errors in our subsequent analysis. After the reduction procedure, one data cube per pointing was produced (i.e. eight, five, and four cubes for the HR-R, HR-I, and LR-Z set-ups, respectively) with a spatial binning of 0.2 arcsec/pix.

\begin{table}[t]
\centering
\caption{Extraction parameters for emission line and continuum maps.}
\label{tab:maps}
\centering
\setlength{\tabcolsep}{3pt}
\begin{tabular}{lcccc}
\hline
\multicolumn{5}{c}{MEGARA}\\
\hline
Line &  $\lambda_c$ (\r{A})  & $\Delta\lambda$ (\r{A}) & $\Delta\lambda_{left}$ (\r{A}) & $\Delta\lambda_{right}$ (\r{A}) \\ 
\hline
 H$\alpha$ & 6563 & 5 & 6596 - 6606 & 6530 - 6540\\
$[SIII]$ & 9532 & 10 & 9472 - 9492 & 9545 - 9565\\
Continuum & 8810 & 5 & - & - \\
\\
\hline
\multicolumn{5}{c}{MUSE}\\
\hline
Line &  $\lambda_c$ (\r{A})  & $\Delta\lambda$ (\r{A}) & $\Delta\lambda_{left}$ (\r{A}) & $\Delta\lambda_{right}$ (\r{A}) \\ 
\hline
 H$\alpha$ & 6563 & 8 & 6531.5 - 6539.5 & 6597 - 6605 \\
 $[SIII]$ & 9069 & 15 & 9010 - 9040 & 9115 - 9145 \\
 Continuum & 8810 & 5 & - & - \\
\hline
\end{tabular}
\tablefoot{
All wavelengths are in the rest frame.
}
\end{table}

Astrometric alignment was achieved by comparing our MEGARA data with those extracted from the publicly available observations of NGC~7742 obtained by the IFS MUSE (Science Verification run on 2014 June 22, ESO Programme \href{http://archive.eso.org/wdb/wdb/eso/sched_rep_arc/query?progid=60.A-9301(A)}{60.A-9301(A)}, M. Sarzi) and already analysed in \cite{ngc7742paper}. The 2D  cross-correlation technique proposed by \citet{1979AJ.....84.1511T} was used to find the spatial shift between MEGARA and MUSE data. We  constructed H$\alpha$ and [SIII]$\lambda$9532 \AA\ emission line maps for the HR-R and LR-Z set-ups, respectively, and a continuum map near the CaT absorption lines for the HR-I set-up. For the emission line maps, a linear behaviour of the continuum emission in the region of interest was assumed choosing appropriate side-bands of a given width for each line. Table \ref{tab:maps} lists in Cols. 1 to 5 (from left to right): the identification of each line; its central wavelength; its width; and the limits of the two continuum side-bands. The upper and lower table panels show these parameters for the  MEGARA and MUSE data, respectively. The followed procedure ensures that our pointings are correctly aligned.

\begin{figure}
    \centering
    \includegraphics[width=\columnwidth]{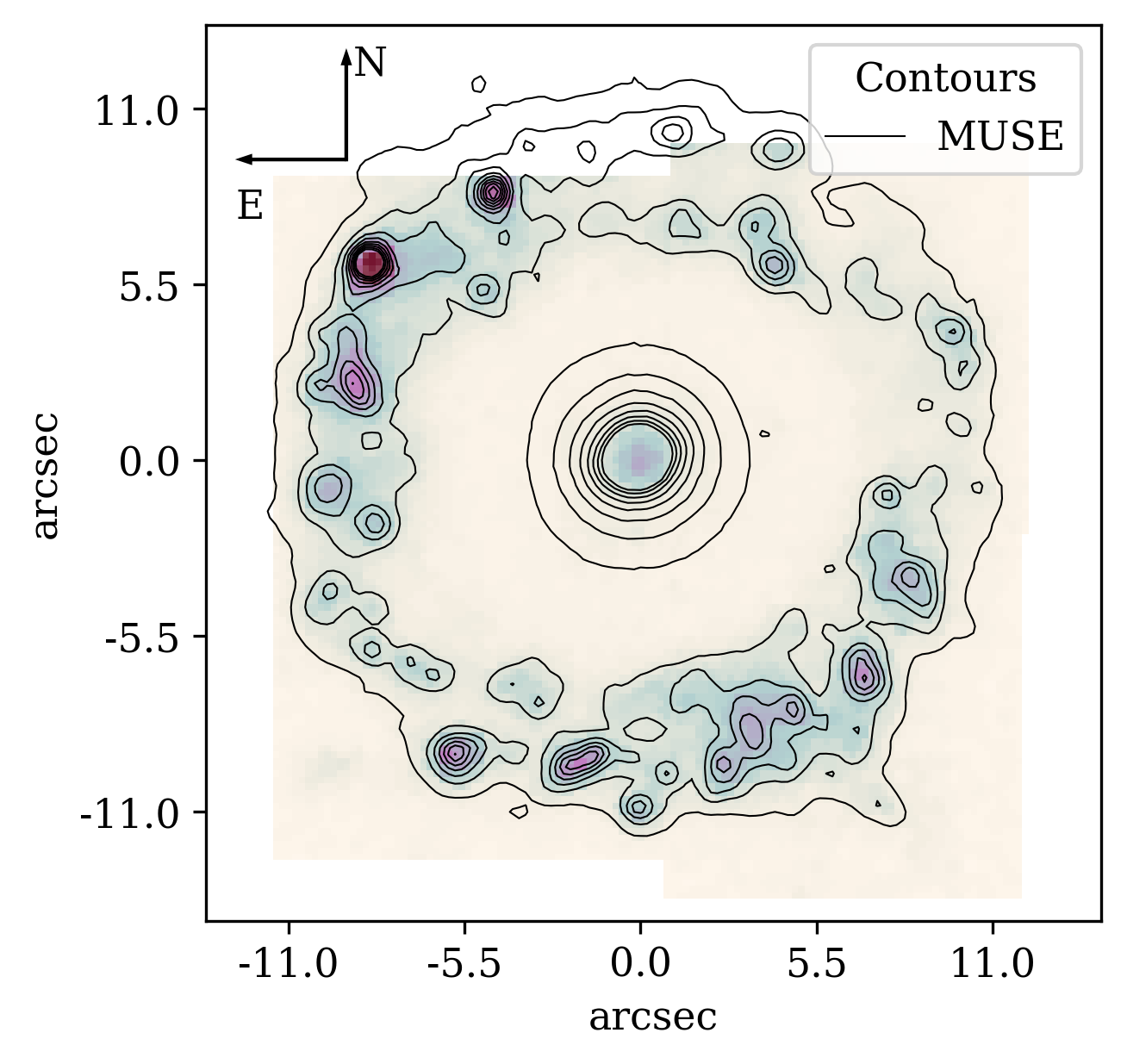}
    \caption{Example of HR-R final data cube compressed in the spectral direction. Superimposed are the contours of the MUSE data (same range of wavelengths). The  colour scale is linear. North is  up and east to the left.}
    \label{fig:megacube}
\end{figure}

Finally, we  combined the overlapping individual cubes into a single cube using the average of the data available in each pixel with no spatial smoothing applied. 
Figure \ref{fig:megacube} shows the resulting cube for HR-R set-up as an example. It is compressed in the spectral direction, and the contours of the data obtained with MUSE showing the consistency with those obtained from MEGARA are superimposed.

\section{Results}

\subsection{HII region selection}

The HII region selection was accomplished as described in the first paper of this series \citep[see][]{ngc7742paper} consisting in the application of an iterative procedure for the detection of high-intensity clumps on an H$\alpha$ emission line map. This method requires as input the maximum and minimum extent of regions (500 pc and the point spread function of observations, respectively) and the relative flux intensity in each of the regions with respect to the emission of its centre (10\%).
Then, the program tried different values for the absolute flux intensity background of the complete map adopting the one that minimises the dispersion of the spatial residuals in
the map.

We consider that this HII region segmentation is appropriate for our MEGARA study since the astrometry correction was made with respect to MUSE data. In addition, since the selected pixel scale for the cube construction is the same as the MUSE scale (0.2 arcsec/pix) we can assure that all the ring HII regions are spatially resolved. Additionally, the measured angular radii of the regions calculated using this method were found to be compatible with radiation bounded ionised nebulae \citep[see ][]{ngc7742paper}. Regions R2, R13, R24, R39, R40, R46, R49, R68, and R82 fall outside the area covered by the FoV of MEGARA, and therefore are not included in the present study. In the end, we   selected a total of 79 HII regions in the ring.

\subsection{Nebular emission line measurements}
\label{sec:line measurements}

We   extracted each region spectrum by integrating the flux inside its corresponding aperture and we   used the code LiMe \citep[LIne MEasuring library][]{2022arXiv221210593F} to separate the different kinematical components present in each emission line. This program uses a defined window for each line and  two additional ones to calculate the continuum below. After the fitting, we took into account only those components that meet the requirement $A_g >3\sigma_l$, with $A_g$ being the Gaussian amplitude and $\sigma_l$ the local standard deviation of the continuum bands selected for each fit. Next, we   measured the radial velocity, $v$, and the velocity dispersion, $\sigma$, of each component in the following emission lines: H$\alpha$, [NII]$\lambda\lambda$6548,84 \AA\ and [SII]$\lambda\lambda$6716,31 \AA\ from the HR-R set-up, and [SIII]$\lambda\lambda$9069,9532 \AA\ from the LR-Z set-up. We   performed an individual fitting for each emission line and adopted a kinematical bond between the components of lines of the same element: [NII]$\lambda$6548 \AA\ with [NII]$\lambda$6584 \AA; [SII]$\lambda$6716 \AA\ with [SII]$\lambda$6731 \AA; and [SIII]$\lambda$9069 \AA\ with [SIII]$\lambda$9532 \AA .

\begin{figure*}
\centering
\includegraphics[width=0.93\textwidth]{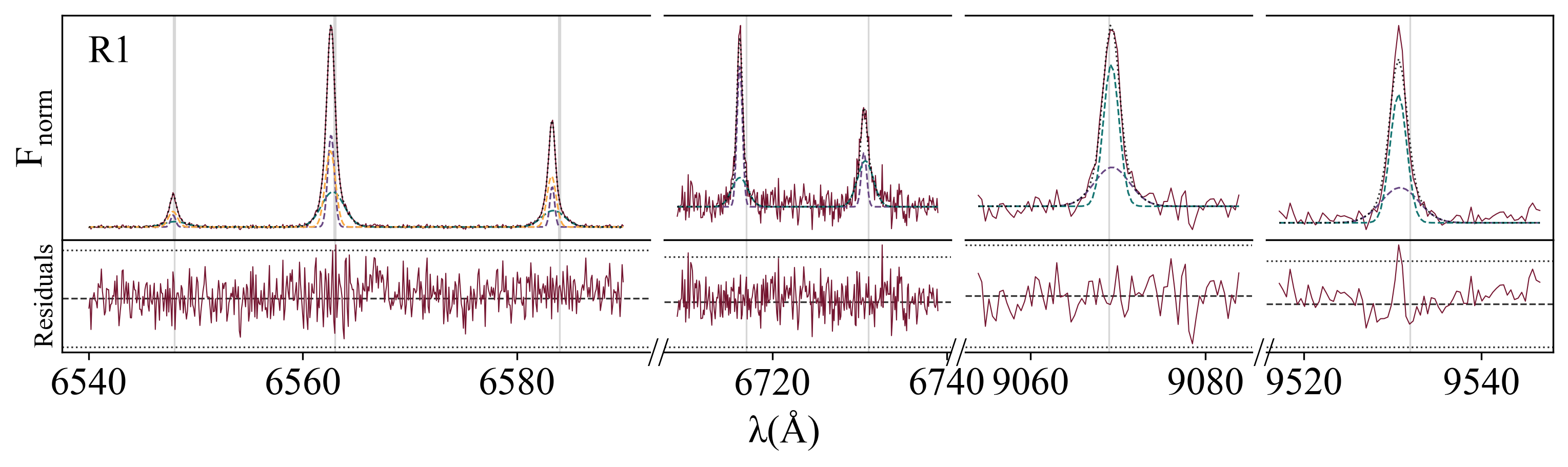}
\includegraphics[width=0.93\textwidth]{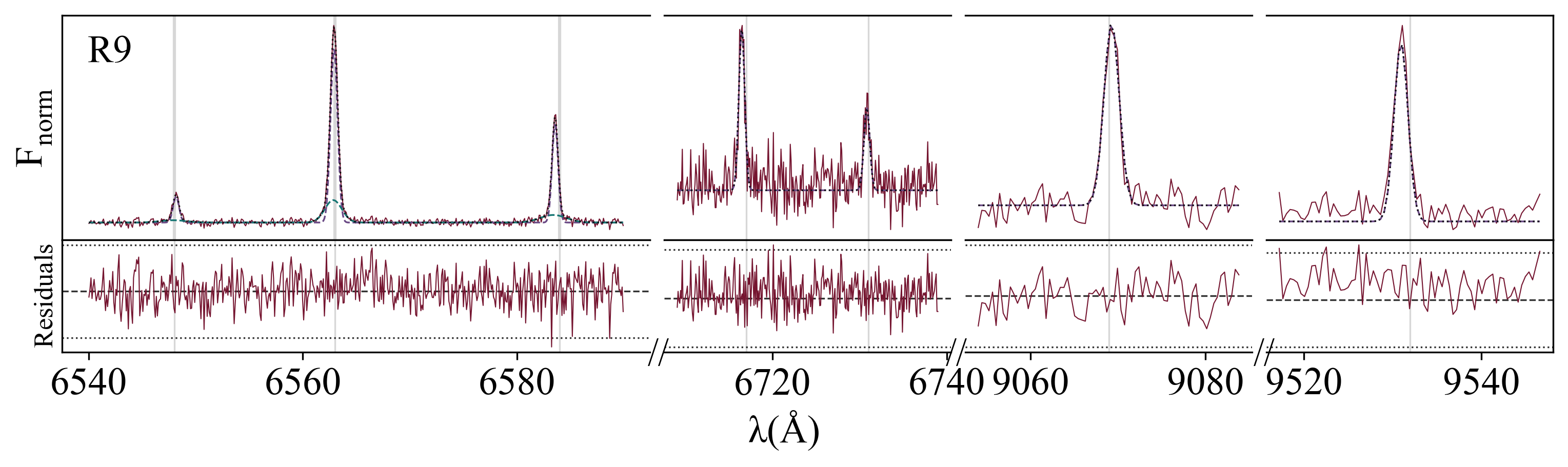}
\caption{Spectra of regions R1 and R9 (top and bottom panels, respectively) showing the kinematical components fitted. From left to right: H$\alpha$ and [NII]$\lambda \lambda$ 6548,6584 \AA ; [SII]$\lambda \lambda $ 6717,6731 \AA ; [SIII]$\lambda$ 9069 \AA ; and [SIII]$\lambda$ 9532 \AA\ emission lines.}
 \label{fig:blend}
\end{figure*}

Figure \ref{fig:blend} shows two examples of the procedure for regions R1 and R9; the panels show (from left to right) the following emission lines: H$\alpha$ and [NII]$\lambda\lambda$6548,84 \AA ; [SII]$\lambda\lambda$6716,31 \AA ; and [SIII]$\lambda$9069 \AA\ and [SIII]$\lambda$9532 \AA. Eight regions out of 79 ($\sim$ 10\%),  labelled R6, R48, R78, R79, R81, R84, R85, and R88, were found to have one only component; the rest   show at least two components in the H$\alpha$ emission line; and only one region, R1, shows two components in the [SIII] lines. In this region, which is shown in the top panel of the figure, three components are clearly identified in the H$\alpha$ and [NII] lines, and also a second can be seen in the[SIII]$\lambda$9532 \AA\ emission line. The region is located near a supernova event: SN 1993R.

\begin{table*}
\centering
\caption{Emission line radial velocities in km/s.}
\label{tab:velocities_MEGARA}
\begin{tabular}{cccccc}
\hline
Region ID      & Component 
& \begin{tabular}[c]{@{}c@{}} v(H$\alpha$) \\ (km/s)\end{tabular}
& \begin{tabular}[c]{@{}c@{}} v([NII]$\lambda$6584 \AA ) \\ (km/s)\end{tabular}
& \begin{tabular}[c]{@{}c@{}} v([SII]$\lambda$6717 \AA ) \\ (km/s)\end{tabular}
& \begin{tabular}[c]{@{}c@{}} v([SIII]$\lambda$9532 \AA) \\ (km/s)\end{tabular}
\\ \hline

R1 & Narrow & 8.45 $\pm$ 0.41 & 8.36 $\pm$ 8.52 & 6.80 $\pm$ 0.65 & 6.71 $\pm$ 1.71 \\
  & Broad & 5.48 $\pm$ 0.85 & 5.76 $\pm$ 12.02 & 7.00 $\pm$ 5.35 & 12.20 $\pm$ 9.46 \\ 
  & Aditional & 15.4 $\pm$ 1.2 & 12.86 $\pm$ 37.74 &  - &  - \\
\\
R3 & Narrow & 16.76 $\pm$ 0.09 & 16.57 $\pm$ 2.90 & 16.23 $\pm$ 0.58 & 15.42 $\pm$ 0.93 \\
  & Broad & 20.70 $\pm$ 0.95 & 18.72 $\pm$ 25.72 & 31.55 $\pm$ 9.52 &  - \\ 
\\
R4 & Narrow & 31.52 $\pm$ 0.20 & 32.54 $\pm$ 2.70 & 30.17 $\pm$ 1.12 & 30.07 $\pm$ 1.29 \\
  & Broad & 36.34 $\pm$ 0.78 & 36.35 $\pm$ 19.61 & 44.89 $\pm$ 5.20 &  - \\ 
  & Aditional & 18.2 $\pm$ 6.8 &  - &  - &  - \\
\hline
\end{tabular}
\tablefoot{
The complete table is available online;   only a part is shown here as an example.
}
\end{table*}

The measured velocities were corrected to rest frame using the redshift derived value of  v = 1652.364 km/s (z = 0.00553). Errors in the velocity of each component are given by the LiMe code using a Monte Carlo procedure and the noise calculated in the continuum bands. Table \ref{tab:velocities_MEGARA}\footnote{The accuracy of the fit and the number of components have been determined by analysing the chi-square of the residuals.} shows the velocity results  in columns 1 to 6: (1) the region ID; (2) the line component; and (3-6) the velocity of the H$\alpha$, [NII], [SII], and [SIII] emission lines, respectively. The corresponding 
errors in the velocity dispersions were calculated as the difference between the FWHM of the largest and smallest fitted Gaussian function taking into account the noise of the fitted spectrum. Using the amplitude of the fitted Gaussian, $A_g$,  and the noise calculated in the residuals of the fit, $\sigma_f$, the error can be calculated as 

\begin{equation}
    \Delta [\sigma] = \frac{\sigma}{2\sqrt{ln(2)}}\cdot ln\left(\frac{1+\sigma_f/A_g}{1-\sigma_f/A_g}\right).
\end{equation}

The measured velocity dispersions of the gas needs to be corrected for instrumental dispersion, $\sigma_{inst}$ and thermal broadening, $\sigma_{th}$, and hence the corrected value for the gas velocity dispersion is 

\begin{equation}
    \sigma_{gas}=\sqrt{\sigma^2_{measured}-\sigma^2_{inst}-\sigma^2_{th}}.
\end{equation}

The instrumental dispersion values are 6.4 km/s and 22.5 km/s for the MEGARA high- and low-resolution configurations, respectively (see Sect. \ref{sec:observations:MEGARA}). The thermal broadening can be calculated assuming a Maxwellian velocity distribution of the ions using the  equation

\begin{equation}
    \sigma _{th}= \sqrt{\frac{k\cdot t_e}{m}},
\end{equation}
where k is the Boltzmann constant (k=1.380649 $\times$ 10$^{-23}$ J/K), t$_e$ is the electron temperature in degrees Kelvin, and m is the mass of the ion  involved (m$_{H^+}$ = 1.6726 $\times$ 10$^{-27}$ kg; m$_{N^+}$ = 2.3257 $\times$ 10$^{-26}$ kg; m$_{S^+}$ = 5.3244 $\times$ 10$^{-26}$ kg; m$_{S^{++}}$ = 5.3243 $\times$ 10$^{-26}$ kg). We   assumed the electron temperature of [SIII] (t$_e$[SIII]) calculated in \citet{ngc7742paper} for each CNSFR;    its average value is 7967 K, and hence the mean thermal corrections are 8.11 km/s for H$^+$; 2.17 km/s for N$^+$; and 1.44 km/s for S$^+$ and S$^{++}$.

\begin{table*}
\centering
\caption{Emission line velocity dispersions in km/s.}
\label{tab:sigma_MEGARA}
\begin{tabular}{cccccc}
\hline
Region ID      & Component & \begin{tabular}[c]{@{}c@{}} $\sigma$(H$\alpha$) \\ (km/s)\end{tabular}
& \begin{tabular}[c]{@{}c@{}} $\sigma$([NII]$\lambda$6584 \AA ) \\ (km/s)\end{tabular}
& \begin{tabular}[c]{@{}c@{}} $\sigma$([SII]$\lambda$6717 \AA ) \\ (km/s)\end{tabular}
& \begin{tabular}[c]{@{}c@{}} $\sigma$([SIII]$\lambda$9532 \AA) \\ (km/s)\end{tabular}
\\ \hline

R1 & Narrow & 10.29 $\pm$ 0.09 & 7.88 $\pm$ 0.15 & 12.41 $\pm$ 0.80 & 18.98 $\pm$ 0.53 \\
  & Broad & 19.60 $\pm$ 0.20 & 18.30 $\pm$ 0.28 & 40.22 $\pm$ 4.14 & 62.64 $\pm$ 6.50 \\ 
  & Aditional & 47.7 $\pm$ 1.1 & 48.30 $\pm$ 2.34 &  - &  - \\
\\
R3 & Narrow & 14.46 $\pm$ 0.08 & 12.67 $\pm$ 0.13 & 13.82 $\pm$ 0.65 & 13.68 $\pm$ 0.44 \\
  & Broad & 43.68 $\pm$ 1.66 & 34.51 $\pm$ 2.02 & 53.56 $\pm$ 8.84 &  - \\ 
\\
R4 & Narrow & 10.53 $\pm$ 0.12 & 12.44 $\pm$ 0.21 & 12.00 $\pm$ 1.34 & 14.06 $\pm$ 0.58 \\
  & Broad & 27.88 $\pm$ 0.59 & 45.04 $\pm$ 3.09 & 33.12 $\pm$ 3.54 &  - \\ 
  & Aditional & 69.2 $\pm$ 10.3 &  - &  - &  - \\
\hline
\end{tabular}
\tablefoot{
The complete table is available online;   only a part is shown here as an example.
}
\end{table*}

Tables \ref{tab:velocities_MEGARA} and \ref{tab:sigma_MEGARA} show the velocity and velocity dispersion results  in  columns 1 to 6: (1) the region ID; (2) the line component; (3-6) the velocity or velocity dispersion of the H$\alpha$, [NII], [SII], and [SIII] emission lines, respectively.  

\subsection{Stellar absorption lines}
\label{EWCaT_MEGARA}

The main objective of our work is to estimate the properties of the stellar population of the clusters associated with the observed CNSFRs. In order to do this, we   used two sets of absorption lines: the MgIb triplet at 5167-5183 detected in the MUSE data (unresolved) and the CaII triplet (CaT) at $\lambda \lambda $ 8498, 8542, 8662 \AA\ (CaT) resolved with MEGARA.

We   measured the equivalent widths (EWs) of these stellar absorption lines by integrating the line fluxes and taking as continuum values the linear interpolation between the mean integrated fluxes in 30 \AA\ continuum bands on both sides of the line. The details of the measuring procedure can be seen in Table 8 in \citet{ngc7469paper}. The EW errors were calculated using the standard deviation in the two continuum bands and propagating them in quadrature. In principle, absorption line EW measurements in objects with different velocity dispersions should be for the broadening of the spectral lines; this effect decreases the continuum level providing lower EW values. However, this correction for MUSE data was found to be negligible (< 1 \AA\ for velocity dispersions of around 170 km/s for MUSE spectral resolution), and so no correction was applied. Table \ref{tab:EW_MEGARA} lists in columns 1  to 3: (1) the region ID; (2) the CaT EW in \AA ; and (3) the MgIb EW also in \AA\ .

\begin{table}
\centering
\caption{Equivalent widths of stellar absorption lines of the observed CNSFRs. }
\label{tab:EW_MEGARA}
\begin{tabular}{ccc}
\hline
Region ID & \begin{tabular}[c]{@{}c@{}} EW(CaII)$^a$ \\ (\AA)\end{tabular} & \begin{tabular}[c]{@{}c@{}} EW(MgI) \\ (\AA)\end{tabular} \\
\hline
R1 & 7.307 $\pm$ 0.264 & 1.686 $\pm$ 0.035  \\ 
R3 & 7.301 $\pm$ 0.278 & 2.329 $\pm$ 0.063  \\ 
R4 & 7.558 $\pm$ 0.303 & 1.868 $\pm$ 0.044  \\ 
R5 & 7.418 $\pm$ 0.338 & 2.147 $\pm$ 0.048  \\ 
R6 & 7.118 $\pm$ 0.318 & 2.141 $\pm$ 0.073  \\ 
R7 & 6.732 $\pm$ 0.339 & 1.972 $\pm$ 0.056  \\ 
R8 & 7.350 $\pm$ 0.299 & 2.156 $\pm$ 0.060  \\ 
R9 & 7.571 $\pm$ 0.346 & 2.013 $\pm$ 0.054  \\ 
R10 & 7.008 $\pm$ 0.297 & 1.987 $\pm$ 0.048  \\ 
R11 & 7.281 $\pm$ 0.211 & 2.840 $\pm$ 0.092  \\ 
\hline
\end{tabular}
\tablefoot{$^a$($\lambda$8542\AA\ + $ \lambda$8662\AA + $ \lambda$8498\AA ). The complete table is available online;  only a part is shown here as an example.
}
\end{table}

The presence of a non-negligible contribution by the nebular continuum can dilute the starlight weakening and distorting the absorption features. This component is usually not taken into account, but it provides information about the origin of the features measured. Thus, we   calculated a dilution factor, D, for each set of lines with respect to a reference average value measured in the spectra of normal spiral galaxy nuclei, which corresponds to old metal-rich stellar populations \citep[EW$_{ref}$(CaII) = 7.7 $\pm$ 0.5 \AA , EW$_{ref}$(MgI) = 5.18 $\pm$ 0.71 \AA ;][]{1990MNRAS.242..271T,1983ApJ...269..466K}. We  calculated this dilution factor as the ratio of the measured EWs to the adopted reference values, D = EW$_{obs}$ / EW$_{ref}$, and compared it to the dilution produced by the nebular continuum associated with a nebula ionised by a young star cluster synthesised using the PopStar code \citep{Popstar} with Salpeter’s IMF \citep[m$_{low}$ = 0.85 M$_\odot$, m$_{up}$ = 120 M$_\odot$][]{Salpeter1955} and an age of 5.5 Ma.

\begin{figure}
\centering
\includegraphics[width=\columnwidth]{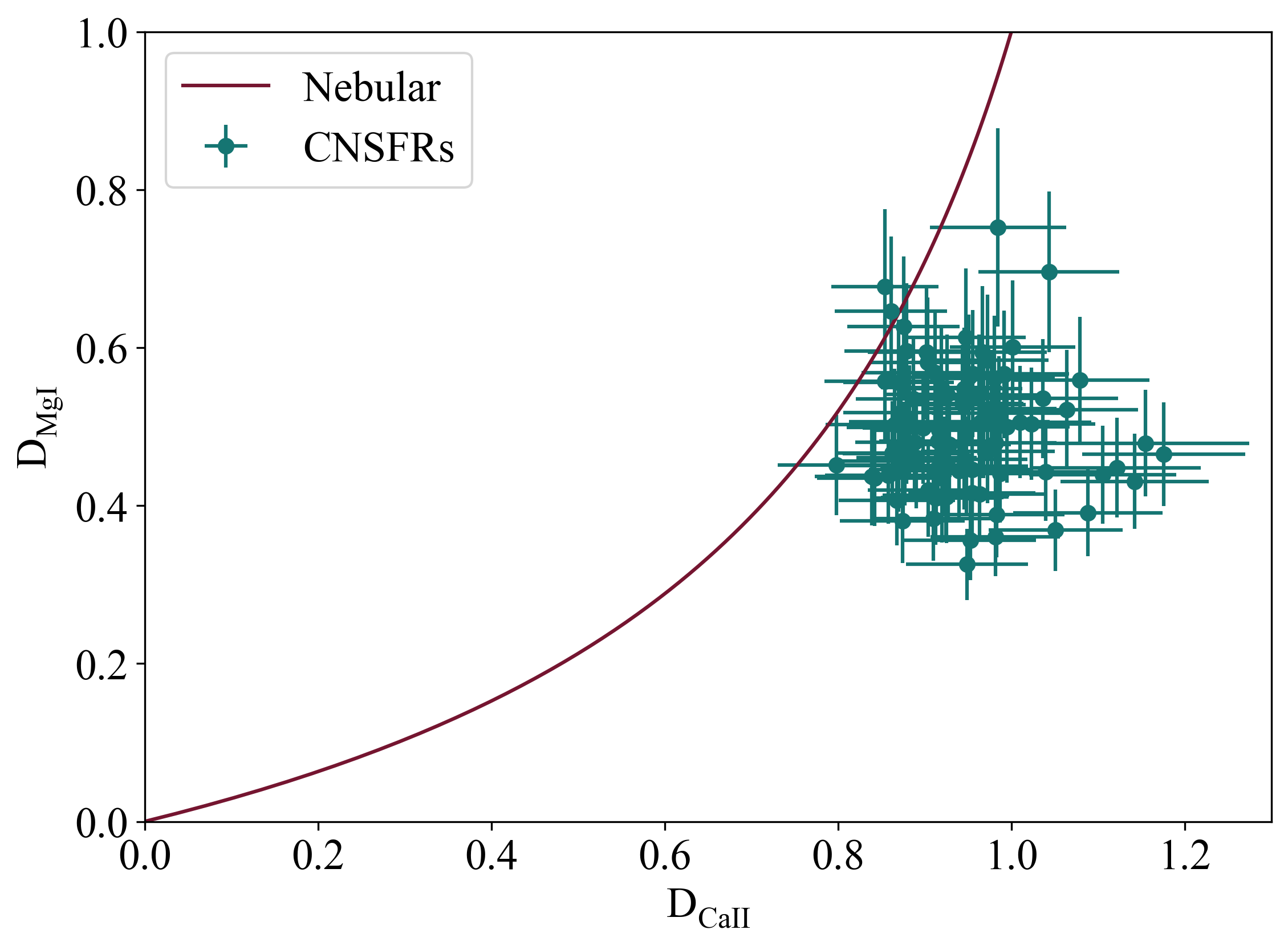}
\caption{
Comparison between the dilution factors calculated for the MgIb and CaT stellar absorption features for the observed CNSFRs. The red lines represent the dilution produced by the nebular continuum in the optical and near-IR.}
 \label{fig:Dilution}
\end{figure}

Figure \ref{fig:Dilution} shows the comparison between the dilution factors, D,  calculated for the MgIb and CaT stellar absorption features for the studied CNSFRs. The red lines represent the dilution produced by the nebular continuum in the optical and near-IR.
We can see that the mean value of the MgIb dilution is about 50\%, consistent with the contribution by a nebular continuum. However, the CaT lines are shown almost undiluted. This result can be explained by the presence of red supergiant stars whose CaT features are stronger than those in normal early spiral galaxies, while the reference value probably consists of a stellar population dominated by red giants. This points to the occurrence of a relatively recent star formation episode. The same behaviour has already been found for circumnuclear star clusters in both rings of NGC~7469 \citep[see][]{ngc7469paper} and in  earlier work \citep{1990MNRAS.242..271T,1993Ap&SS.205...85G,1995A&A...301...55O}.

\subsection{Stellar velocity dispersions}

Stellar velocity dispersions can be obtained from the stellar absorption lines. In the second paper of this series \citep[see ][]{ngc7469paper} we used the CaT lines detected in the MUSE data for this calculation. However, the spectral resolution reached in the MUSE data was insufficient to resolve the CaT lines in the case of the HII ring regions in NGC~7742, and therefore we acquired higher resolution data using MEGARA, which provides about  14 km/s of spectral resolution (see Sect. \ref{sec:observations:MEGARA}). As in the case of NGC~7469, the clusters studied in this work have red supergiant stars dominating the near-IR range of the spectrum and the CaT $\lambda \lambda $ 8498, 8542, 8662 \AA\  lines (see Sect. \ref{EWCaT_MEGARA}) are easily detected.
The use of the near-IR spectra provides two principal advantages for this study: there is little     contamination by TiO bands and nebular lines,   and the velocity resolution at longer wavelengths is higher than in the blue part of the spectrum for the same spectral dispersion. Additionally, using the cross-correlation technique, stellar weak lines present in this part of the spectra also contribute to the goodness of the results. 

We   used the cross-correlation technique proposed by \citet{1979AJ.....84.1511T}. In this traditional method, a comparison between a stellar template with the observed stellar population spectrum is done. The assumption  made assumes that a galaxy spectrum can be represented by the sum of different stellar spectra with different velocity offsets convolved with a broadening function: $g\sim \alpha [t(n)\ast b(\lambda-\delta)]$, where \textit{b} is the broadening function; \textit{g} and \textit{t} are the cluster and the stellar template spectra, respectively; and $\delta$ is the velocity shift between the stars. The width of this broadening function is calculated using two correlations: the cluster spectrum with the stellar template \textit{(g $\otimes$ t)} and the stellar template with itself \textit{(t $\otimes$ t)}. The convolution products are applied assuming a periodic spectrum with discrete Fourier transforms, and the broadening function is assumed to be Gaussian.
However, this last assumption is not correct for the \textit{(t $\otimes$ t)} correlation at spectral resolutions as high as the one provided by  MEGARA,  and this technique needs to be modified. We  used the adapted methods proposed in \cite{2023arXiv231004133Z} of the \textit{(g $\otimes$ g)} correlation, which is broader than the other two  and whose Gaussian behaviour can be ensured. In order to test these two additional methods and the traditional ones with our data, we  selected a region with high S/N (R52, located in the pointing with the higher exposure time in the HR-I set-up). For the stellar template, we   used late-type red giants and supergiants obtained from the MEGARA stellar library \citep{2020MNRAS.493..871G} aligning all available stellar spectra in velocity and calculating the average between them, verifying that no apparent broadening is introduced in the procedure. The use of only two stellar types of stars for the correlation analysis could introduce errors in our velocity dispersion measures. However, the use of CaT lines minimises the mismatch between the stellar template and the cluster spectrum since they are very strong in most stars of moderate to high metallicity. 


\begin{figure*}
\centering
\includegraphics[width=\linewidth]{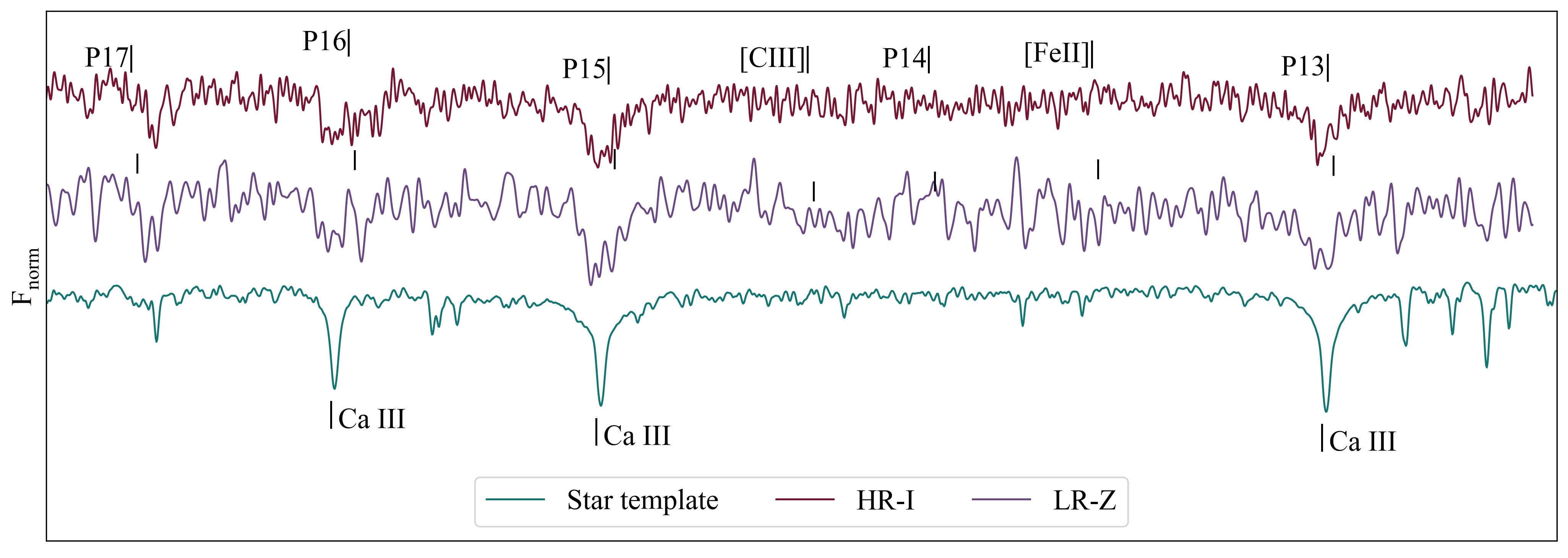}
\includegraphics[width=\linewidth]{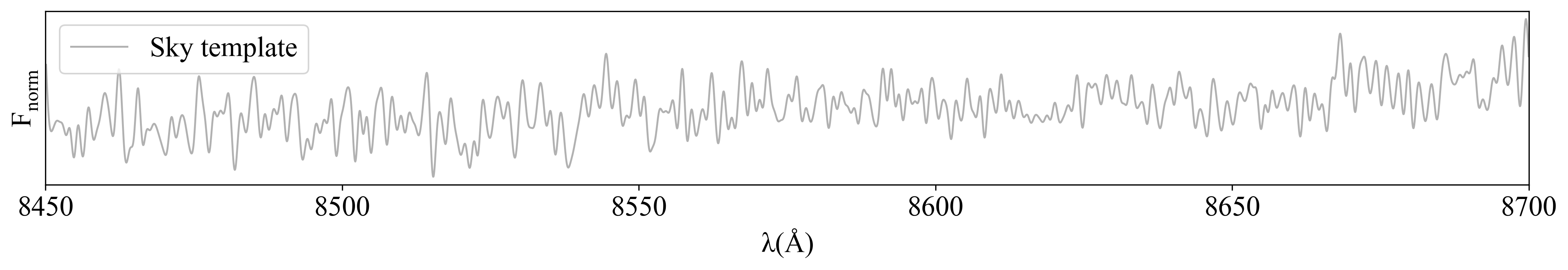}
\caption{Typical spectra used to measure the stellar velocity dispersions. Top panel: R52 absorption spectrum from MEGARA/HR-I and LR-Z configurations shown as red and purple lines, respectively. The absorption stellar lines and emission lines present in this wavelength range are indicated. The stellar template is also shown as a blue line. Bottom panel: Sky template (see text).}
 \label{fig:Spectrum_plot_HR-I}
\end{figure*}

The top panel of Fig. \ref{fig:Spectrum_plot_HR-I} shows the stellar spectrum of the selected  cluster R52 at high and low MEGARA spectral resolutions, with red and purple lines respectively. The stellar template is   shown with a blue line and the nebular lines present in this spectral range are marked. We can see that the S/N values of these spectra are low, even in this region that was observed with twice the average exposure time. This could be due to the noise component or to errors associated with the subtraction of the sky background, which is very prominent in the near-IR. Thus, we  calculated a mean spectrum of the sky, using the MEGARA fibres earmarked for this purpose. The lower panel of the figure shows this spectrum, which  shares some features with the cluster spectrum.

We   used the modified version of the  Tonry \& Davis method described above for the application of the cross-correlation technique to the data obtained from the MEGARA in the HR-I configuration in the spectral range  8450 \AA to 8850 \AA. Prior to this, we   binned the spectrum into 4069 logarithmic wavelength bins corresponding to a velocity resolution of $\sim$ 3.4 km/s. Then, we   estimated the continuum by fitting a second-order polynomial after masking the features present in the spectrum. The masks were built assuming a width of 3 \AA\ and 10 \AA\ on each side of the central wavelength of the nebular and stellar lines involved, respectively. This continuum was subsequently   subtracted.

\begin{figure}
\centering
\includegraphics[width=\linewidth]{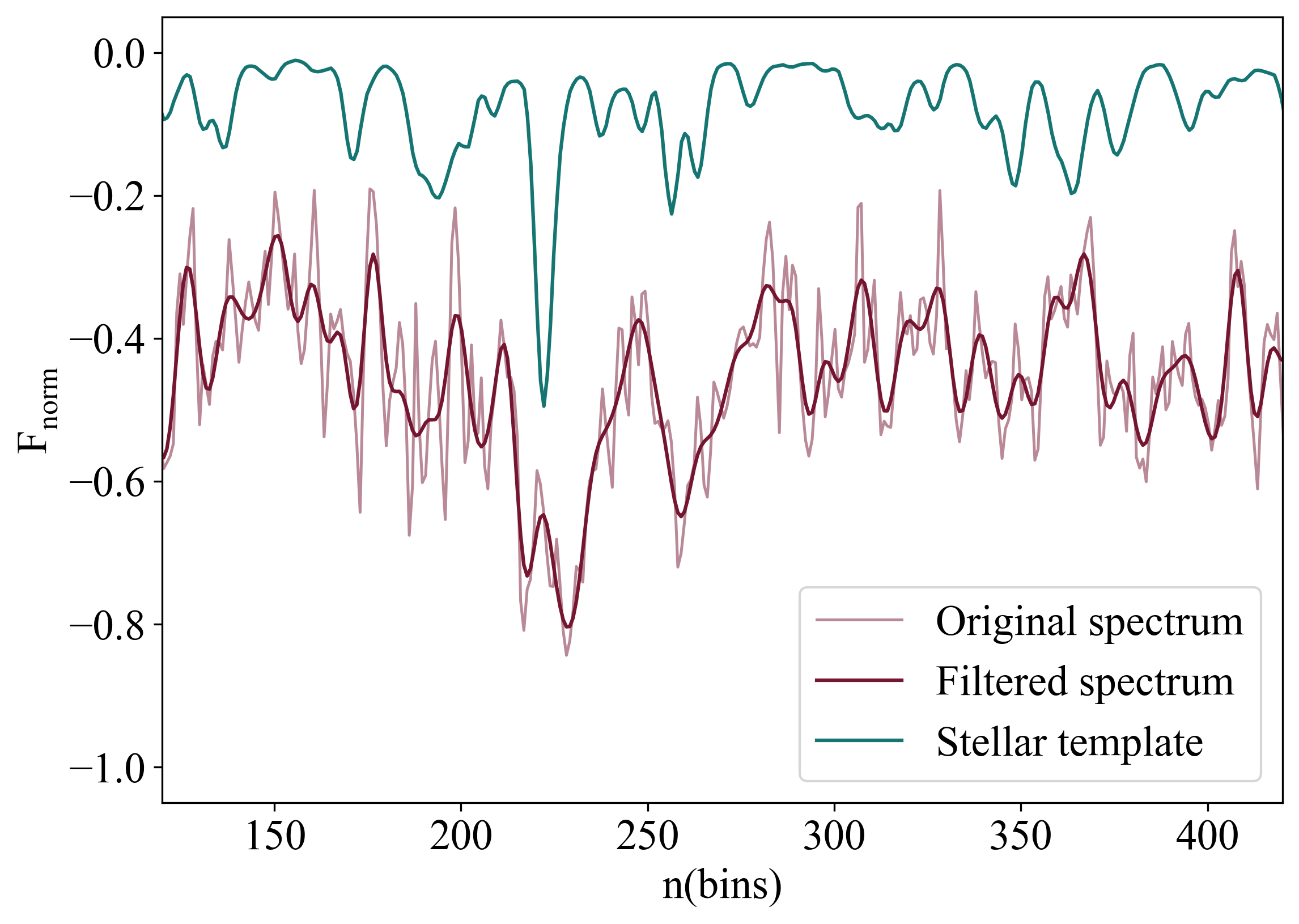}
\caption{Absorption line spectrum of R52 from the MEGARA/HR-I configuration, in logarithmic wavelength. The light and dark red lines show the spectrum before and after applying the pass-band filter for higher frequencies. The stellar template is also shown (blue line).}
 \label{fig:Filtered_R52}
\end{figure}

Next, the high- and low-frequency variations were filtered using a band-pass filter in the Fourier  spectrum transform. The minimum and maximum wave numbers are k$_{min}$ = 3 and k$_{max}$ = 350, which correspond to wavelength values lower than 10 \AA\ and to the nominal MEGARA/HR-I spectral resolution. The low-frequency variations are associated with continuum subtracted errors and the high-frequency variations   with the noise component. Figure \ref{fig:Filtered_R52} shows, in a small range of wavelengths, in light red the original spectrum and in dark red the spectrum after the band-pass filter. The stellar template is also shown in blue for comparison. We can see that this filter removes part of the noise, recovering some stellar features. However, the S/N continues to be low and we can see residuals, for example in the central part of the strongest line. Thus, the noise present in our spectrum probably comes from the errors associated with the pipeline sky subtraction. Therefore, we   decided to use the data without the sky subtraction to be able to quantify this additional component in the cross-correlation technique and to obtain a more reliable result in the stellar dispersion determination.

An important step in the cross-correlation procedure is the subtraction of the non-stellar component of the spectra, which can introduce errors in the velocity dispersions of up to 5 km/s \citep[see ][]{2023arXiv231004133Z}. In order to remove correctly the nebular component of the spectrum, we   first removed the proper motion of the cluster by shifting the spectrum to the rest-frame position. For this step, we   used the position of the \textit{(g $\otimes$ t)} cross-correlation. Then, we   masked the nebular lines shown in Fig. \ref{fig:Spectrum_plot_HR-I} assuming a width of 0.4 \AA\ at each side of their central wavelength.

\begin{figure}
\centering
\includegraphics[width=0.98\columnwidth]{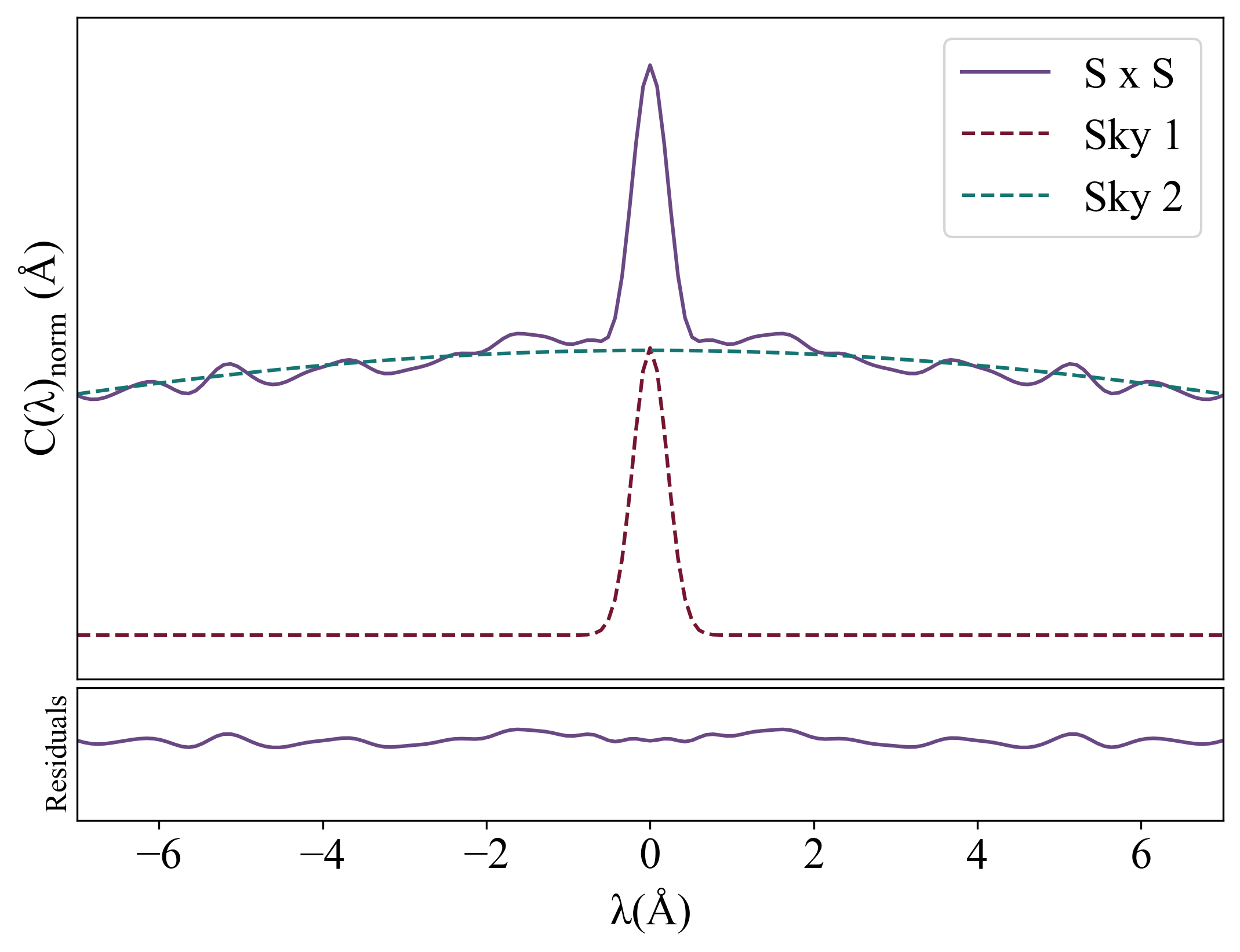}
\includegraphics[width=0.98\columnwidth]{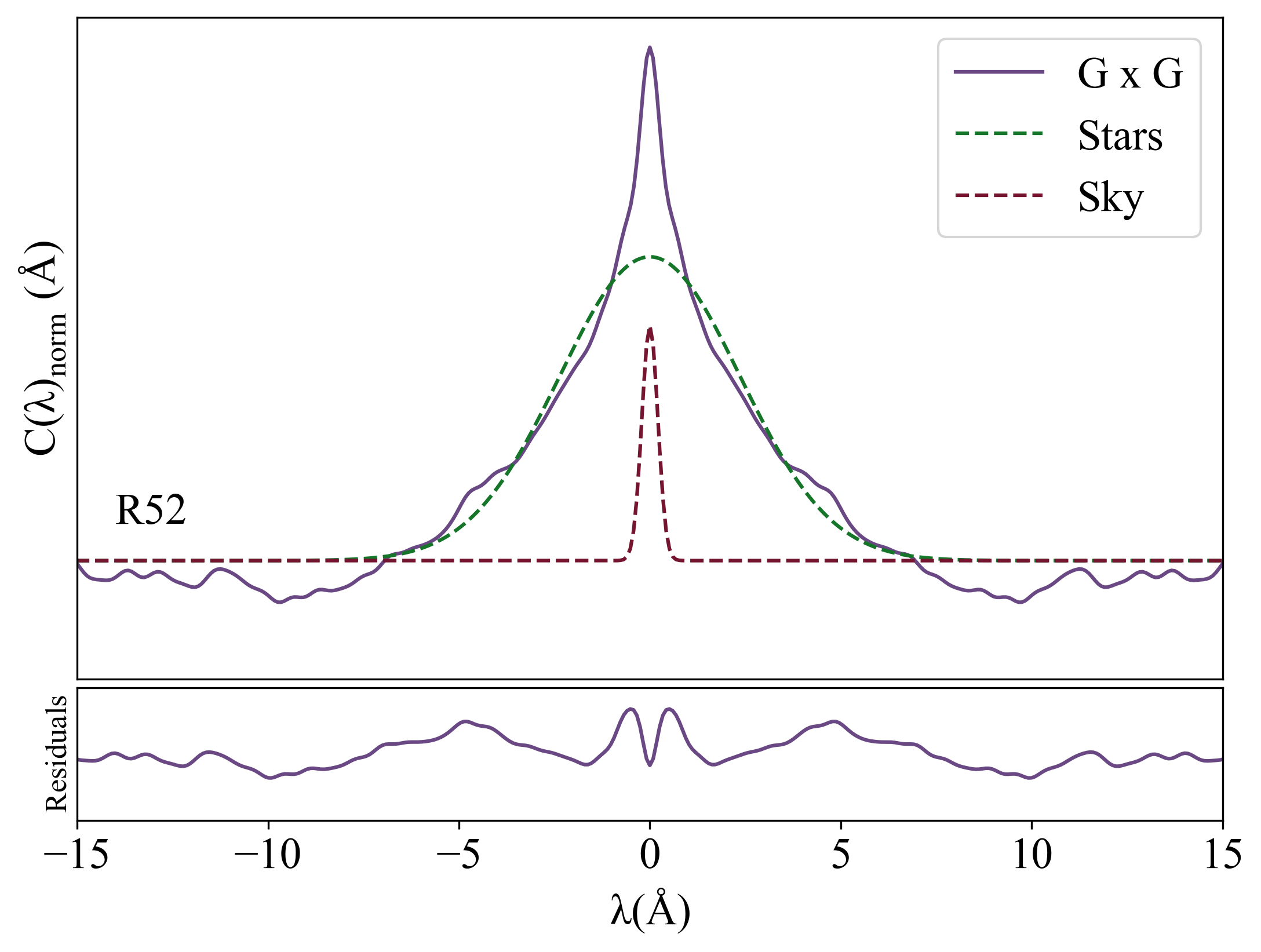}
\includegraphics[width=0.98\columnwidth]{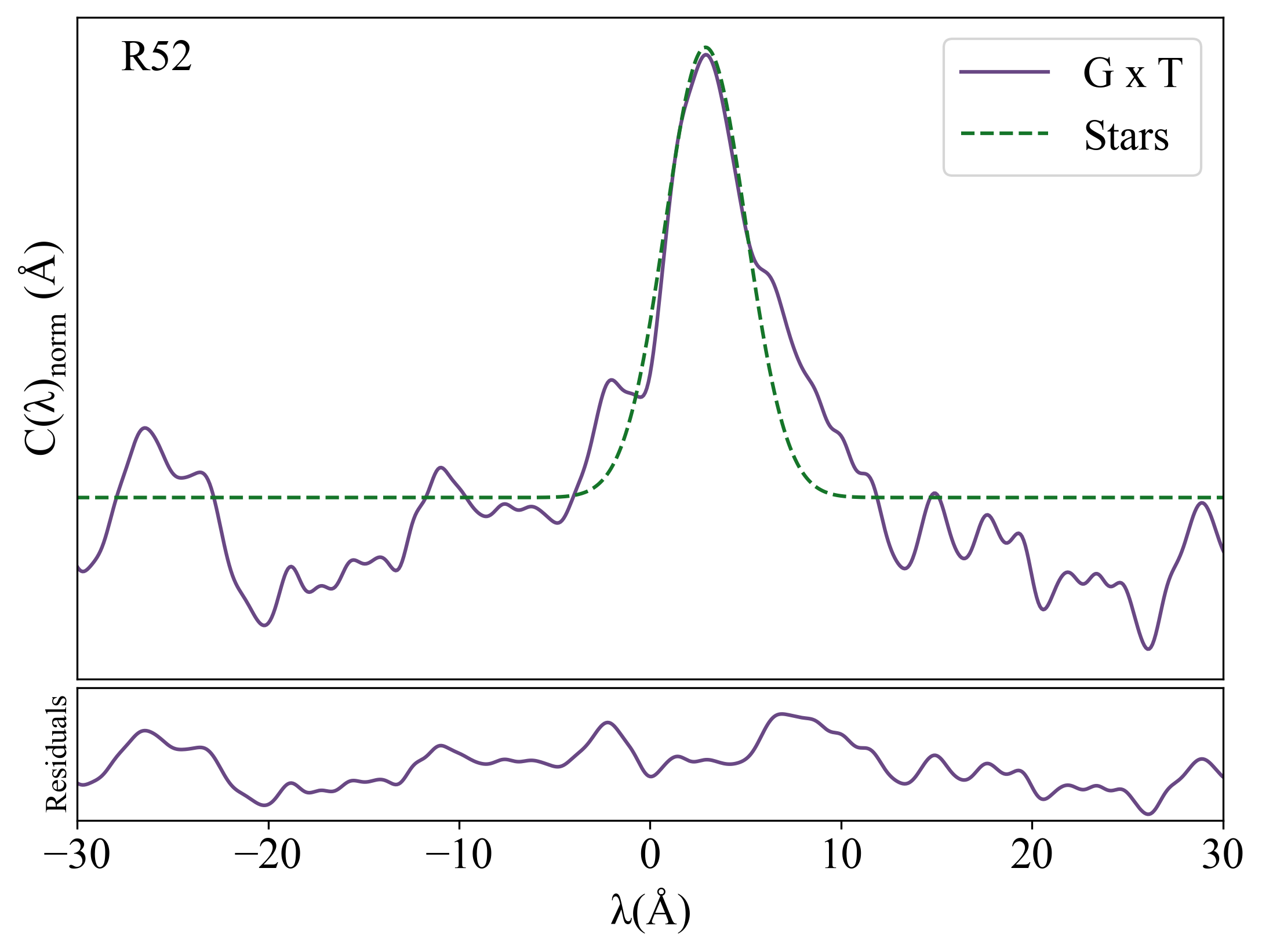}
\caption{Example of auto-cross-correlations calculated in this work. Top panel:  Auto-cross-correlation function of the sky template. Middle panel: Auto-cross-correlation of the spectrum of region R52. Bottom panel: Cross-correlation between the cluster spectrum and the stellar template. The lower panel of each plot shows the residuals of the fitting.}
 \label{fig:example_corr}
\end{figure}

Finally, we   calculated four cross-correlation functions for this HII region, \textit{(t $\otimes$ t)}, \textit{(g $\otimes$ t)}, \textit{(g $\otimes$ g)},  and \textit{(s $\otimes$ s)}, with \textit{t} being the stellar template, \textit{g} the cluster spectrum, and \textit{s} the mean sky spectrum of our observations. The top panel of Fig. \ref{fig:example_corr} shows the \textit{(s $\otimes$ s)} correlation. Two components can be seen, one associated with the sky lines and another associated with the detector response. They were fitted with two Gaussian functions and their widths are $\mu_{ss1}$ = 0.2098 \AA\ and $\mu_{ss2}$ = 12.1445 \AA, respectively. The width of the first component corresponds to a value of $\Delta \lambda$ = 0.15 \AA\ in $\sigma$-units, which is slightly lower than the MEGARA resolution for this set-up, $\Delta \lambda$ = 0.18.

The top panel of Fig. \ref{fig:example_corr} shows, as an example, the \textit{(g $\otimes$ g)} correlation for the R52 cluster. In this figure we can identify the component associated with the sky lines and the component associated with the stars of the cluster. According to this, we   fitted  simultaneously two Gaussian functions, one of them using the $\mu_{ss1}$ value calculated for the \textit{(s $\otimes$ s)} correlation. The width of the cluster component is $\mu_{gg}$ = 2.2680 \AA. The correlation (g $\otimes$ t) is shown in the lower panel of the same figure. In this case the sky emission affects only the cluster spectrum, and hence the sky component does not show up. The cluster component is $\mu_{gt}$ = 2.1147 \AA.

We   combined these results, together with the width of the \textit{(t $\otimes$ t)} correlation function ($\mu_{tt}$ = 1.241 \AA) and we   applied the three methods studied in \cite{2023arXiv231004133Z}:
\[ (i) \quad \sigma = \sqrt{\frac{\mu_{gg}^2-\mu_{tt}^2}{2}},\]
\[ (ii) \quad \sigma = \sqrt{\mu_{gg}^2-\mu_{gt}^2} ,\]
\[ (iii) \quad \sigma = \sqrt{\mu_{gt}^2-\mu_{tt}^2} .\]
The last equation is the one proposed by \citet{1979AJ.....84.1511T}. We   performed these calculations for R52, correcting methods (i) and (ii) for the non-Gaussian behaviour of the (t $\otimes$ t) correlation. The final results are (i) $\sigma$ = 27.87 km/s; (ii) $\sigma$ = 28.42 km/s; and (iii) $\sigma$ = 28.44 km/s. We can see that the three methods show very similar results, and thus we can conclude that the reduction and treatment of these high-resolution data and the application of the cross-correlation method is correct.

\begin{table}
\centering
\caption{Stellar cluster velocity dispersions in units of km/s.}
\label{tab:sigma_Stars_MEGARA}
\begin{tabular}{cc}
\hline
Region ID &  \begin{tabular}[c]{@{}c@{}} $\sigma_*$(CaT) \\ (km/s)\end{tabular}\\
\hline
R1 & 22.61 $\pm$ 6.65 \\
R3 & 31.43 $\pm$ 4.68 \\
R5 & 77.08 $\pm$ 1.37 \\
R7 & 20.19 $\pm$ 4.39 \\
R8 & 37.75 $\pm$ 19.32 \\
R14 & 19.34 $\pm$ 5.87 \\
R15 & 27.88 $\pm$ 6.61 \\
R16 & 50.82 $\pm$ 3.95 \\
R22 & 37.10 $\pm$ 9.98 \\
R23 & 38.75 $\pm$ 8.66 \\
R26 & 19.26 $\pm$ 6.86 \\
R31 & 45.87 $\pm$ 9.38 \\
R33 & 47.73 $\pm$ 5.41 \\
R36 & 72.04 $\pm$ 0.49 \\
R41 & 50.82 $\pm$ 4.95 \\
R45 & 56.26 $\pm$ 14.67 \\
R47 & 38.95 $\pm$ 13.01 \\
R50 & 44.64 $\pm$ 4.94 \\
R51 & 36.78 $\pm$ 6.83 \\
R52 & 28.40 $\pm$ 9.85 \\
R53 & 50.53 $\pm$ 1.13 \\
R55 & 44.67 $\pm$ 7.63 \\
R57 & 14.46 $\pm$ 4.49 \\
R60 & 33.36 $\pm$ 7.53 \\
R63 & 32.93 $\pm$ 7.31 \\
R69 & 14.25 $\pm$ 6.79 \\
R72 & 26.40 $\pm$ 4.24 \\
R74 & 23.56 $\pm$ 5.80 \\
R75 & 43.76 $\pm$ 6.35 \\
R88 & 20.65 $\pm$ 3.52 \\
\hline
\end{tabular}
\end{table}

Therefore, for the rest of the CNSFRs, we   decided to apply method (iii) in order not to use the \textit{(g $\otimes$ g)} correlation, due to the sky subtraction problems. The velocity dispersion errors were   calculated as the semi-difference between the largest and smallest Gaussian width that can be fitted considering the asymmetries in the correlation peak, as suggested in \citet{2023arXiv231004133Z}, although it could lead to the error being somewhat overestimated.  
Applying the described methodology, we   obtained reliable CaT velocity dispersion values for 30 of the selected CNSFR regions. They are shown in Table \ref{tab:sigma_Stars_MEGARA}, and they take values from 14.3 to 77.1 km/s.

\section{Discussion}

\subsection{Cluster sizes}
\begin{figure}
\centering
\includegraphics[width=\columnwidth]{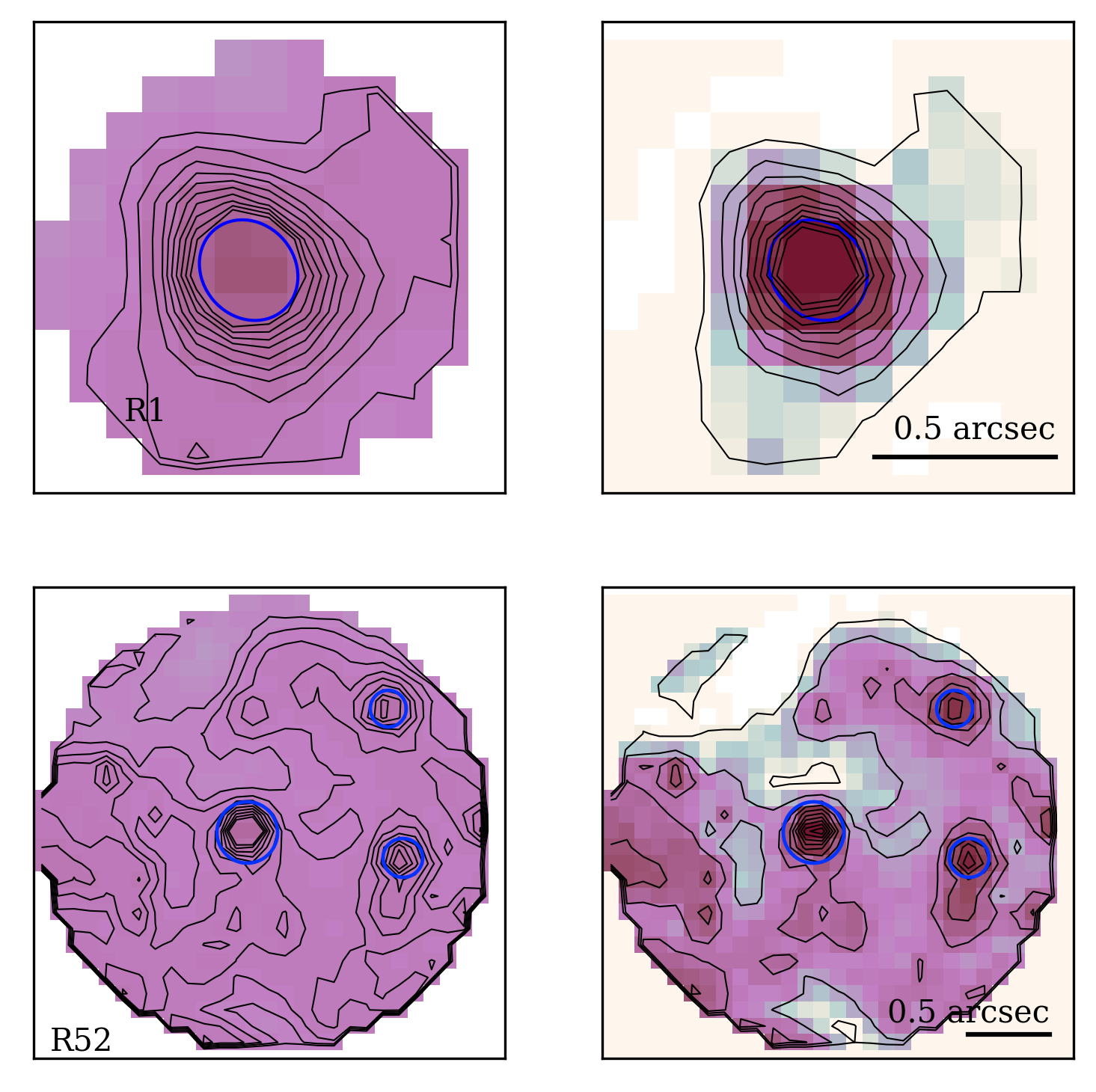}
\caption{Results of the stellar cluster radius measurement procedure for regions R1 and R52, upper and lower panels, respectively (see text for details). The angular scale is given in the  corner of each panel.}
 \label{fig:plot_radios_example}
\end{figure}

We   calculated the effective radii of the star clusters from the \textit{Hubble} Space Telescope (HST)  continuum image. These data were acquired on 2009 November 5 with the Wide Field and Planetary Camera 2 (WFPC2) and the F555W continuum filter where stellar clusters are easily identified. These data are a part of proposal \href{https://archive.stsci.edu/proposal_search.php?mission=hst&id=6276}{6276} (IP: J. Westphal) organised in three exposures of 480 s. The data were downloaded from the \textit{Hubble} Legacy Archive and  reduced  by the Space Telescope Science Institute (STScI) using the available calibration files taken for this observation and keeping in mind different dithering positions.

The effective radii of the star clusters were calculated using the following procedure. First, we   estimated the region background by fitting a three-order polynomial. After subtracting it, we fitted the surface brightness of each knot present in the observed clusters assuming a two-dimensional Gaussian profile. Figure \ref{fig:plot_radios_example} shows two examples of the described procedure as applied to regions R1, and R52. The left panels of the figure show the F555W WFPC2-HST image for each selected region and the right panels show the same image after background subtraction. Selected clusters are shown by blue circles. The radius of each knot is taken as 1/2$\cdot$ FWHM. There are three regions, R72, R75, and R88, where no knot is readily appreciable in the continuum at 5550 \AA, and thus we did not calculate their sizes.

As opposed to other CNSFRs from the literature, this galaxy does not seem to have large star-forming complexes, and  $\sim$ 60 \% of the CNSFRs (16 out of 27) seem to host single clusters and $\sim$ 27 \% (7 out of 27) are composed  of two clusters. Three regions show three different knots, and  one shows four. We can compare these values with the results obtained for NGC~7469 \citep[see][]{ngc7469paper}, which shows two regions with respectively 20 and 19 knots,  and only 15 \%  of the regions (4 of 27) seem to have only one star cluster. These last results, are similar to those found for the regions analysed in NGC~2903 showing more than 20 knots \citep{2009MNRAS.396.2295H}, NGC~3310  hosting two great complexes with 28 and 30 clusters each \citep{2010MNRAS.402.1005H}, and  NGC~3351 with a large region showing 17 distinct knots \citep{2007MNRAS.378..163H}. In the case of NGC~7469, we imagine  this effect to be due to the galaxy being  further away since in this case we may be selecting larger star-forming complexes, due to lack of spatial resolution; however, the distances to the other three galaxies mentioned are of the same order as the one studied here. In particular,
NGC~3310 at a similar distance to NGC~7742 seems to have only one region hosting an individual knot. It is probable that the difference between the CNSFRs is related to their different formation processes   which in the latter case seem to be due to  gas accretion after a minor merger event.


\begin{figure}
\centering
\includegraphics[width=\columnwidth]{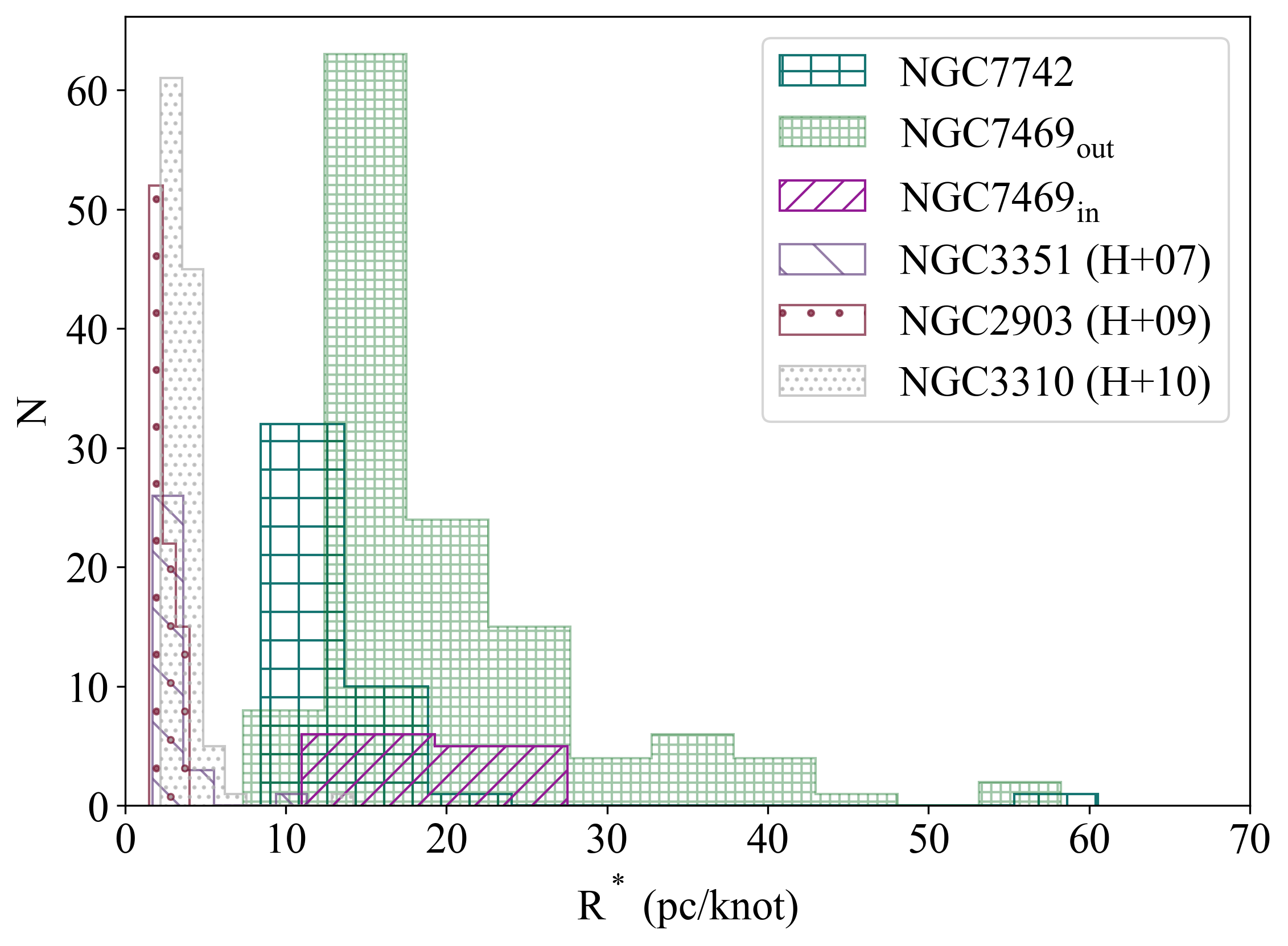}
\caption{Histograms of the distributions of the radius of each knot in the CNSFRs of NGC~7742  compared to those using data from the literature. }
 \label{fig:plot_radios_stars}
\end{figure}

The effective radii of the studied single knots range between 8.5 and 60.5 pc, with a mean value of 12.6 pc. Figure \ref{fig:plot_radios_stars} shows the distribution of the radius of each knot for the CNSFRs in this work and those found in the literature used for comparison \citep{2009MNRAS.396.2295H,2007MNRAS.378..163H,2010MNRAS.402.1005H}. The range of radius found for NGC~7469 is wider than in the other galaxies, and the same is true for the mean radii for the inner and outer ring regions. On the other hand, NGC 3351, NGC 2903, and NGC 3310 have smaller knots than NGC 7742.

\subsection{Dynamical masses} \label{sec:Dynamical_masses:MEGARA}

We   calculated the dynamical masses for each observed cluster from its measured CaT velocity dispersion and size assuming a virialised system and using the expression given by \citep{1996ApJ...466L..83H, 1996ApJ...472..600H}
\begin{equation}
M^{dyn}=3\cdot \sigma^2 \cdot \frac{R}{G} ,   
\end{equation}
where $\sigma$ is the one-dimensional velocity dispersion of the system, R is its radius, and G is the gravitational constant.

\begin{figure}
\centering
\includegraphics[width=\columnwidth]{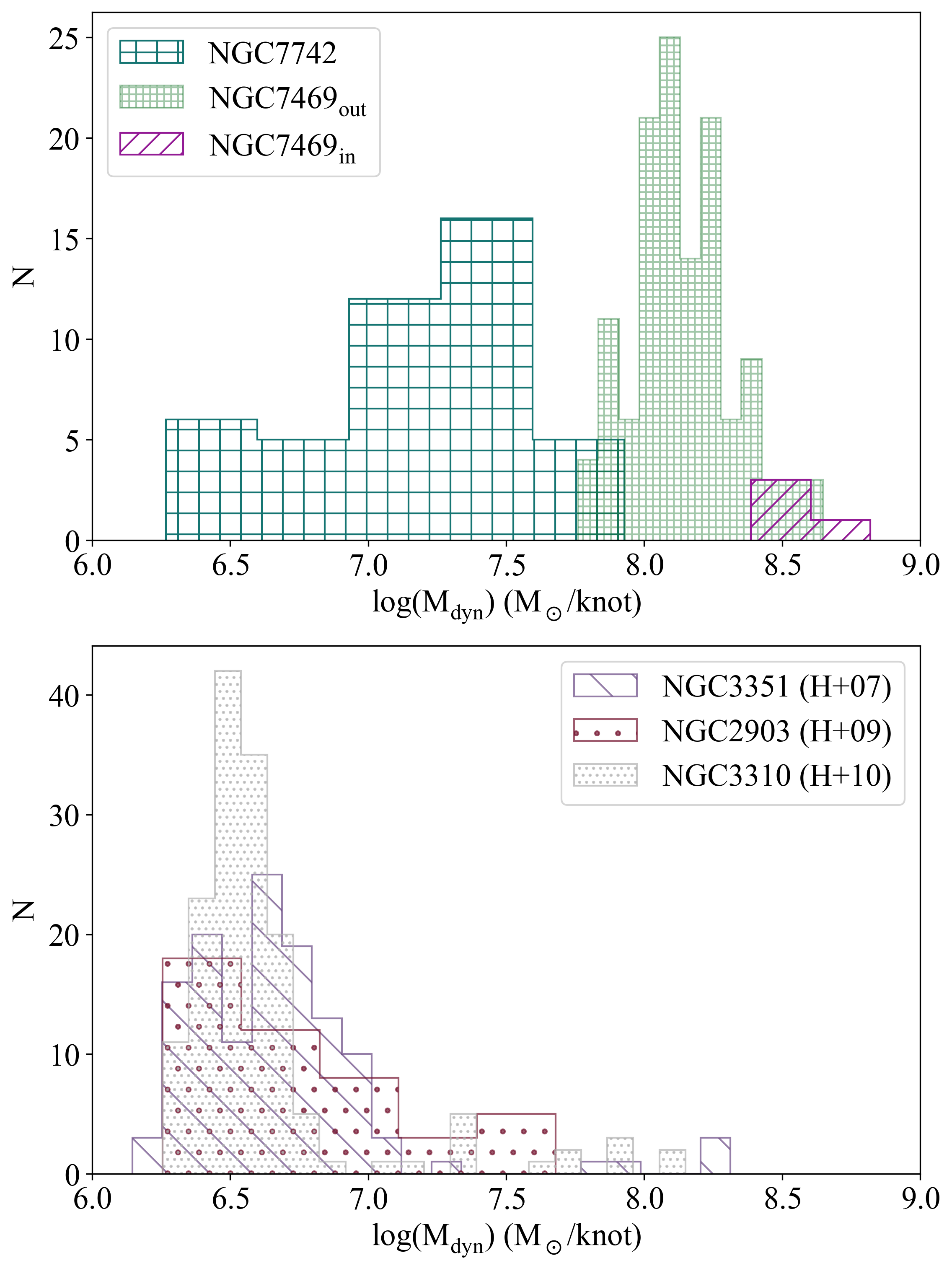}
\caption{Distributions of the dynamical masses of each knot for the CNSFRs. The top panel shows the results in this work and the outer and inner ring regions of NGC~7469. The bottom panel shows data from the literature \citep[][labelled (H+07), (H+09), and (H+10) respectively]{2010MNRAS.402.1005H,2007MNRAS.378..163H,2009MNRAS.396.2295H}.}
 \label{fig:plot_masses_knots}
\end{figure}

This calculation assumes that the clusters (i) have spherical symmetry; (ii) are gravitationally bound; and (iii) have an isotropic velocity distribution.
 Most of the clusters show only one knot, and  thus the calculated masses under this assumption seem to be appropriate. However, a small number are clearly multiple; in this case the total mass of the cluster was estimated by adding the masses of its individual knots inferred using the velocity dispersion of the entire region a procedure that  most likely result in an overestimate. 
 
 Figure \ref{fig:plot_masses_knots} shows the distribution of the cluster dynamical masses for NGC~7742 compared with those of the inner and outer rings of NGC~7469. We can see that NGC~7742 knots show lower masses, which is to be expected from their smaller sizes. However, they tend to be more massive than the CNSFRs from the literature \citep[][NGC~3351, NGC~2903, and NGC~3310]{2010MNRAS.402.1005H,2007MNRAS.378..163H,2009MNRAS.396.2295H}.

\begin{table}
\centering
\setlength{\tabcolsep}{3pt}
\caption{Stellar cluster masses and sizes for observed CNSFRs.}
\label{tab:mdyn_MEGARA}
\begin{tabular}{ccccc}
\hline
\begin{tabular}[c]{@{}c@{}}Region \\ ID\end{tabular} & N$_{knots}$ & \begin{tabular}[c]{@{}c@{}}R$^* _{mean}$ \\ (pc/knot)\end{tabular} 
& \begin{tabular}[c]{@{}c@{}} M$_{dyn}$ \\ (M$_\odot$)\end{tabular} & \begin{tabular}[c]{@{}c@{}} M$_{ion}$/M$_{dyn}$ \\ (per cent)\end{tabular} \\
\hline

R1 & 1 & 13.9 & (6.2 $\pm$ 3.6) $\times$ 10$^{6}$ & 2.57\\ 
R3 & 2 & 15.0 & (25.9 $\pm$ 2.7) $\times$ 10$^{6}$ & 1.47\\ 
R5 & 1 & 16.3 & (84.6 $\pm$ 3.0) $\times$ 10$^{6}$ & 0.15\\ 
R7 & 1 & 9.9 & (3.5 $\pm$ 1.5) $\times$ 10$^{6}$ & 6.61\\ 
R8 & 1 & 20.1 & (2.5 $\pm$ 2.6) $\times$ 10$^{7}$ & 1.32\\ 
R14 & 2 & 12.8 & (8.3 $\pm$ 2.0) $\times$ 10$^{6}$ & 2.20\\ 
R15 & 1 & 12.1 & (8.2 $\pm$ 3.9) $\times$ 10$^{6}$ & 1.45\\ 
R16 & 1 & 10.1 & (22.8 $\pm$ 3.6) $\times$ 10$^{6}$ & 2.26\\ 
R22 & 1 & 10.1 & (12.2 $\pm$ 6.5) $\times$ 10$^{6}$ & 2.24\\ 
R23 & 2 & 14.6 & (38.3 $\pm$ 8.6) $\times$ 10$^{6}$ & 0.73\\ 
R26 & 2 & 14.6 & (9.5 $\pm$ 2.3) $\times$ 10$^{6}$ & 3.00\\ 
R31 & 4 & 10.1 & (74.4 $\pm$ 7.6) $\times$ 10$^{6}$ & 0.40\\ 
R33 & 1 & 15.2 & (30.2 $\pm$ 6.8) $\times$ 10$^{6}$ & 0.77\\ 
R36 & 1 & 10.6 & (481.8 $\pm$ 6.6) $\times$ 10$^{5}$ & 0.77\\ 
R41 & 3 & 14.8 & (100.4 $\pm$ 6.7) $\times$ 10$^{6}$ & 0.59\\ 
R45 & 2 & 18.1 & (10.0 $\pm$ 2.3) $\times$ 10$^{7}$ & 0.19\\ 
R47 & 2 & 12.2 & (3.2 $\pm$ 1.1) $\times$ 10$^{7}$ & 0.55\\ 
R50 & 1 & 10.1 & (17.6 $\pm$ 3.9) $\times$ 10$^{6}$ & 1.21\\ 
R51 & 1 & 71.3 & (8.4 $\pm$ 3.1) $\times$ 10$^{7}$ & 0.39\\ 
R52 & 3 & 14.9 & (31.4 $\pm$ 4.9) $\times$ 10$^{6}$ & 1.46\\ 
R53 & 1 & 10.1 & (22.6 $\pm$ 1.0) $\times$ 10$^{6}$ & 0.58\\ 
R55 & 3 & 14.1 & (73.4 $\pm$ 6.7) $\times$ 10$^{6}$ & 0.13\\ 
R57 & 2 & 14.7 & (5.4 $\pm$ 1.1) $\times$ 10$^{6}$ & 7.07\\ 
R60 & 1 & 13.1 & (12.7 $\pm$ 5.8) $\times$ 10$^{6}$ & 0.90\\ 
R63 & 1 & 10.1 & (9.6 $\pm$ 4.3) $\times$ 10$^{6}$ & 1.33\\ 
R69 & 1 & 14.4 & (2.5 $\pm$ 2.4) $\times$ 10$^{6}$ & 4.16\\ 
R74 & 1 & 19.2 & (9.3 $\pm$ 4.6) $\times$ 10$^{6}$ & 1.14\\ 
\hline
\end{tabular}
\end{table}

Table \ref{tab:mdyn_MEGARA}  lists for each region in columns 1 to 5: (1) the region ID; (2) the number of stellar knots present in the region; (3) the mean radius in pc; (4) the dynamical mass  in M$_{\odot}$; and (5) the ratio of ionising to dynamical mass. To calculate this ratio, we   used the ionising masses calculated in the first paper of the series from the number of hydrogen ionising photons, Q(H$\alpha$), and the equivalent width of the H$\beta$ emission line to take into account the cluster evolution. We used a Salpeter IMF $\phi (m) = m^{-\alpha}$, with $\alpha = 2.35$, $m_{low}(M_\odot)$ = 0.85, and $m_{up}(M_\odot)$ = 120, which seems the most suitable for the age of our regions.

\begin{figure}
\centering
\includegraphics[width=\columnwidth]{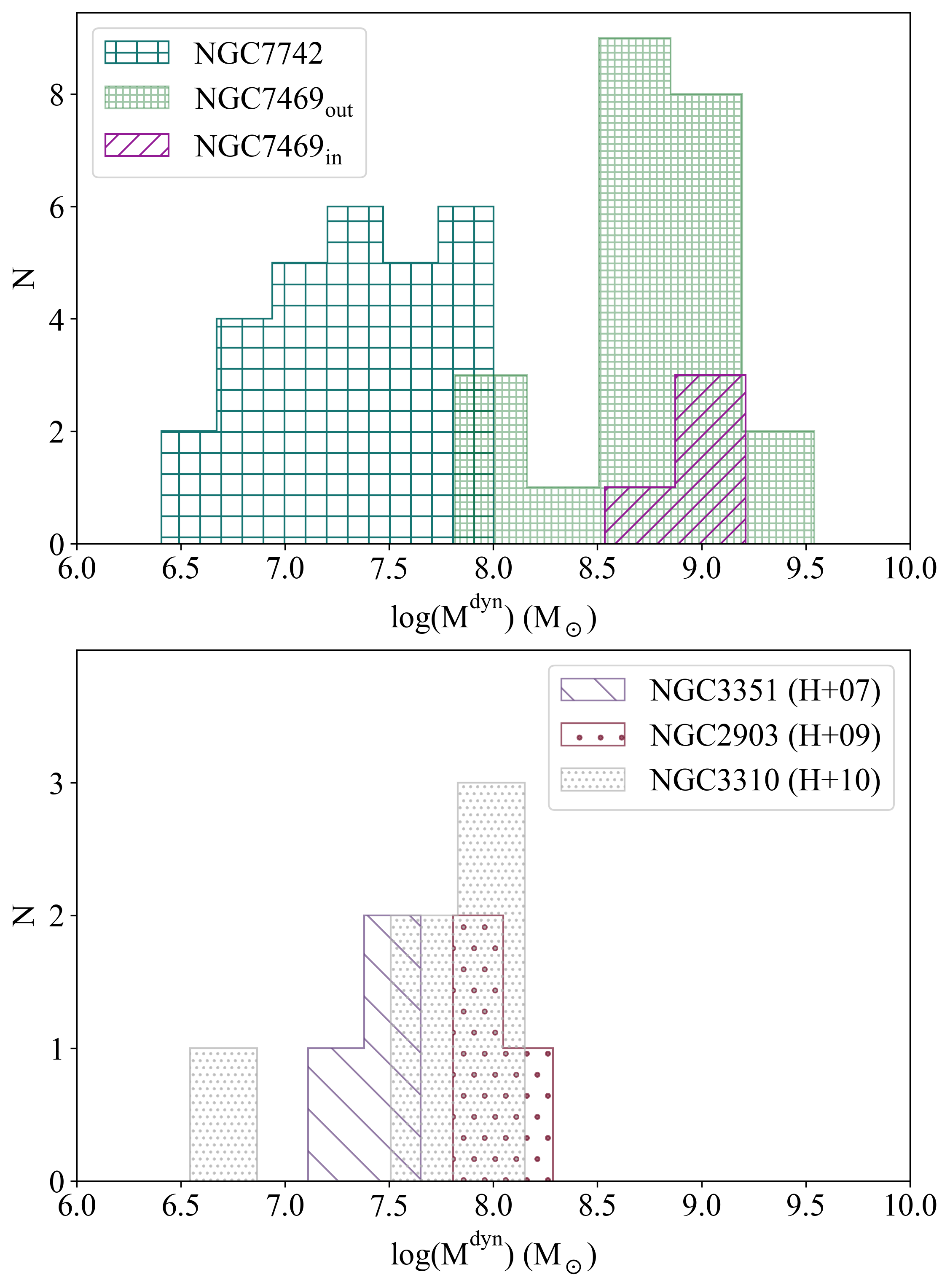}
\caption{Histograms of the distribution of dynamical masses for the CNSFRs in this work, the outer and inner ring regions of NGC~7469, and data from the literature as labelled \citep[][(H+07), (H+09), and (H+10) respectively]{2010MNRAS.402.1005H,2007MNRAS.378..163H,2009MNRAS.396.2295H}.}
 \label{fig:hist_masa_din}
\end{figure}

Figure \ref{fig:hist_masa_din} shows the distribution of the calculated dynamical masses for each CNSFR. The dynamical masses of inner and outer ring regions of  NGC~7469, calculated in the second paper of this series are also shown. Additionally, we show the masses of CNSFRs reported in the literature for NGC~3351, NGC~2903, and NGC~3310, which were calculated with the same method described in this work \citep{2010MNRAS.402.1005H,2007MNRAS.378..163H,2009MNRAS.396.2295H}. The dynamical masses of NGC~7742 clusters range from 2.5 $\times$ 10$^6$  to 1.0 $\times$ 10$^8$ M$_\odot$, values similar to those found in the rest of the galaxies shown, but lower than those obtained for the two rings of NGC~7469. 

The ratio of ionising to dynamical masses has values ranging from 0.15 to 7.07 \%, implying a contribution by recent star formation for the ring regions larger than found in other galaxies. For comparison, this percentage ranges from 0.02 to 0.91 \%  for the CNSFRs in the outer ring of NGC~7469 whose distance to the galactic nucleus is similar to that of the NGC~7742 ring to its nucleus.

\subsection{Kinematics of stars and ionised gas}

\begin{figure*}
\centering
\includegraphics[width=\textwidth]{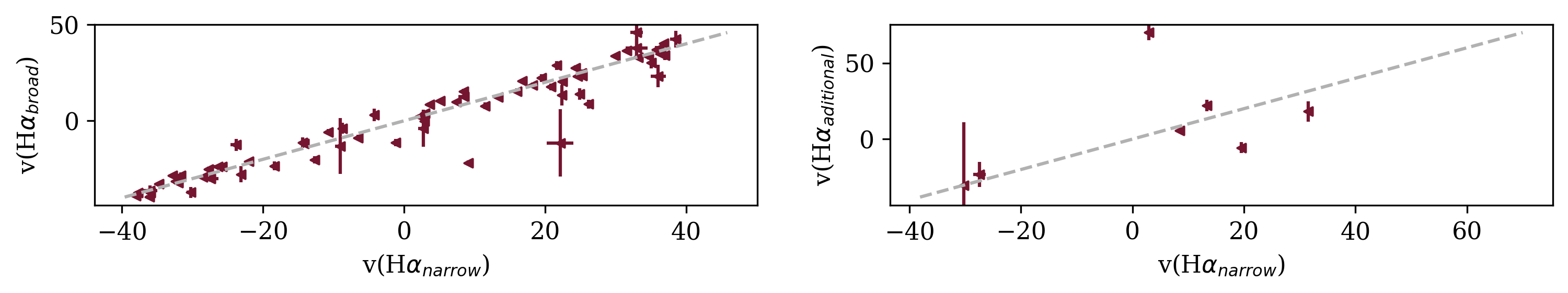}
\caption{Velocity of different components of H$\alpha$ measured in each region. The narrow component velocity is shown against the broad and additional components, in the left and right panels, respectively. All velocities are shown in km/s.}
 \label{fig:Ha_nebular_lines}
\end{figure*}

In Sect. \ref{sec:line measurements} we   derive the radial velocities and velocity dispersions of the ionised gas in the apertures of the selected HII regions. Our results suggest the existence of at least two kinematically distinct H$\alpha$ components that can be identified in the spectra of CNSFRs with  high S/N. We   compared their radial velocities in order to find out which of them are associated with the ionising clusters.

The left panel of Fig. \ref{fig:Ha_nebular_lines} shows the comparison of the velocities of the H$\alpha$ broad and narrow components with the one-to-one line represented by a dashed grey line. We can see that the two radial velocities coincide within the errors.  Only two regions seem to fall off the  one-to-one line: R62, and R65. An additional component of 
the H$\alpha$ emission line has been identified in seven of the regions, and a comparison of their velocities and those of the narrow component is shown in the right panel of the same figure. Only three of these additional components, in regions R1, R7, and R67, seem to be associated with the ionising clusters, and hence they can be attributed to local effects inside the regions. 
In the other four cases, that component may have an external origin.  

\begin{figure*}
\centering
\includegraphics[width=\textwidth]{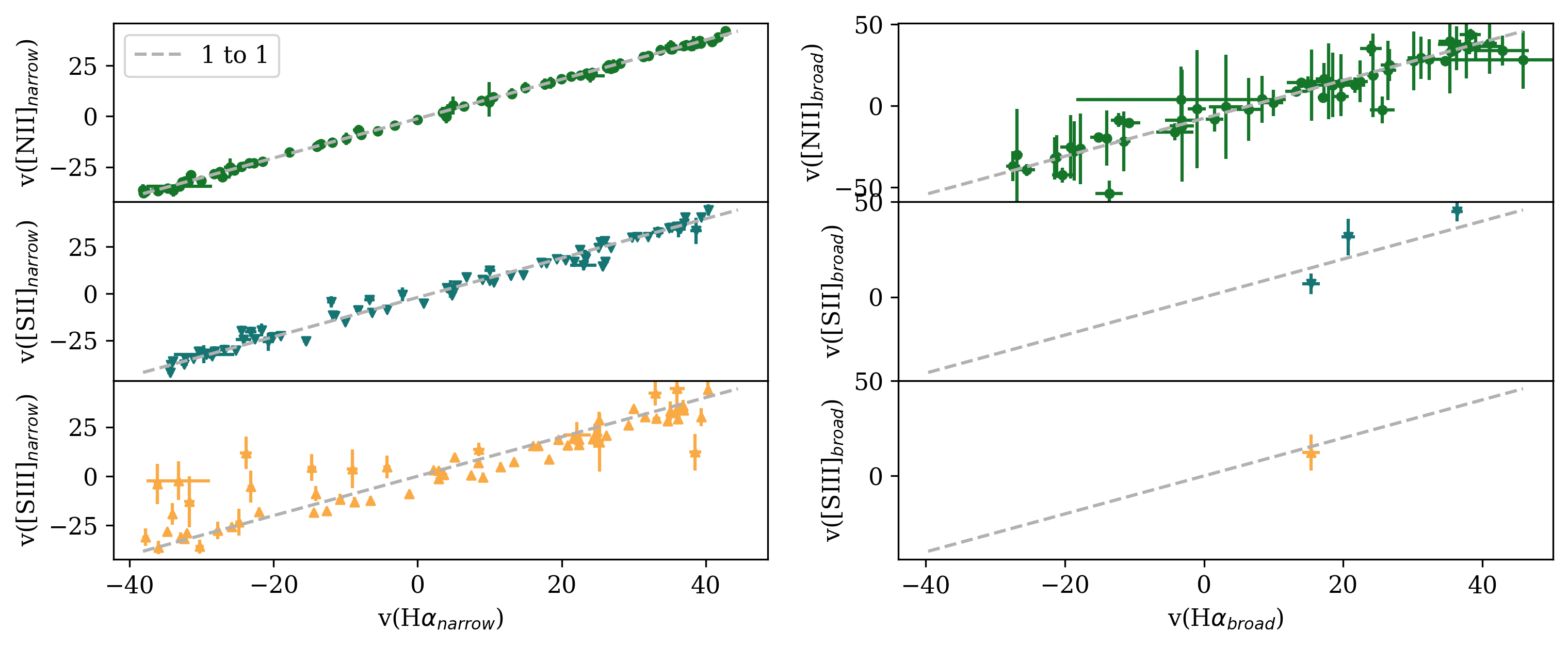}
\caption{Comparison of the radial velocity of the different measured components of the [NII], [SII], and [SIII] lines in each region (from top to bottom) with those of the narrow and broad H$\alpha$  components (left and right panels). All velocities are shown in km/s.}
 \label{fig:nebular_lines}
\end{figure*}

The left panels of Fig. \ref{fig:nebular_lines} show the gas velocity of the narrow component of forbidden emission lines as a function of the corresponding H$\alpha$ component. Results for the [NII]$\lambda $6584 \AA , [SII]$\lambda $6731 \AA , and [SIII]$\lambda$9532 \AA\ emission lines are shown from top to bottom. All regions follow the one-to-one relation within the errors, hence assuring that the selected kinematical components are associated with the observed HII regions. Only in the case of the [SIII] emission line do a few regions seem to deviate from the identity relation; these regions show large observational errors related to low S/N spectra. The right panels of the figure show the same relations, but for the broad-line component of both forbidden and H$\alpha$ lines. All the [NII] broad components follow the one-to-one relation; their association with the observed HII regions is guaranteed. For the [SII] lines, only in three of the regions is the presence of a broad component   consistent with the S/N, and only one region shows a broad component in the [SIII] emission lines; this region, R1, displays  complex kinematics with three different width velocity components, clearly identified in the H$\alpha$ and [NII]$\lambda\lambda$6548,84 \AA\ lines, that seem to be associated with the HII region (see Fig. \ref{fig:blend}).


\begin{figure*}
\centering
\includegraphics[width=\columnwidth]{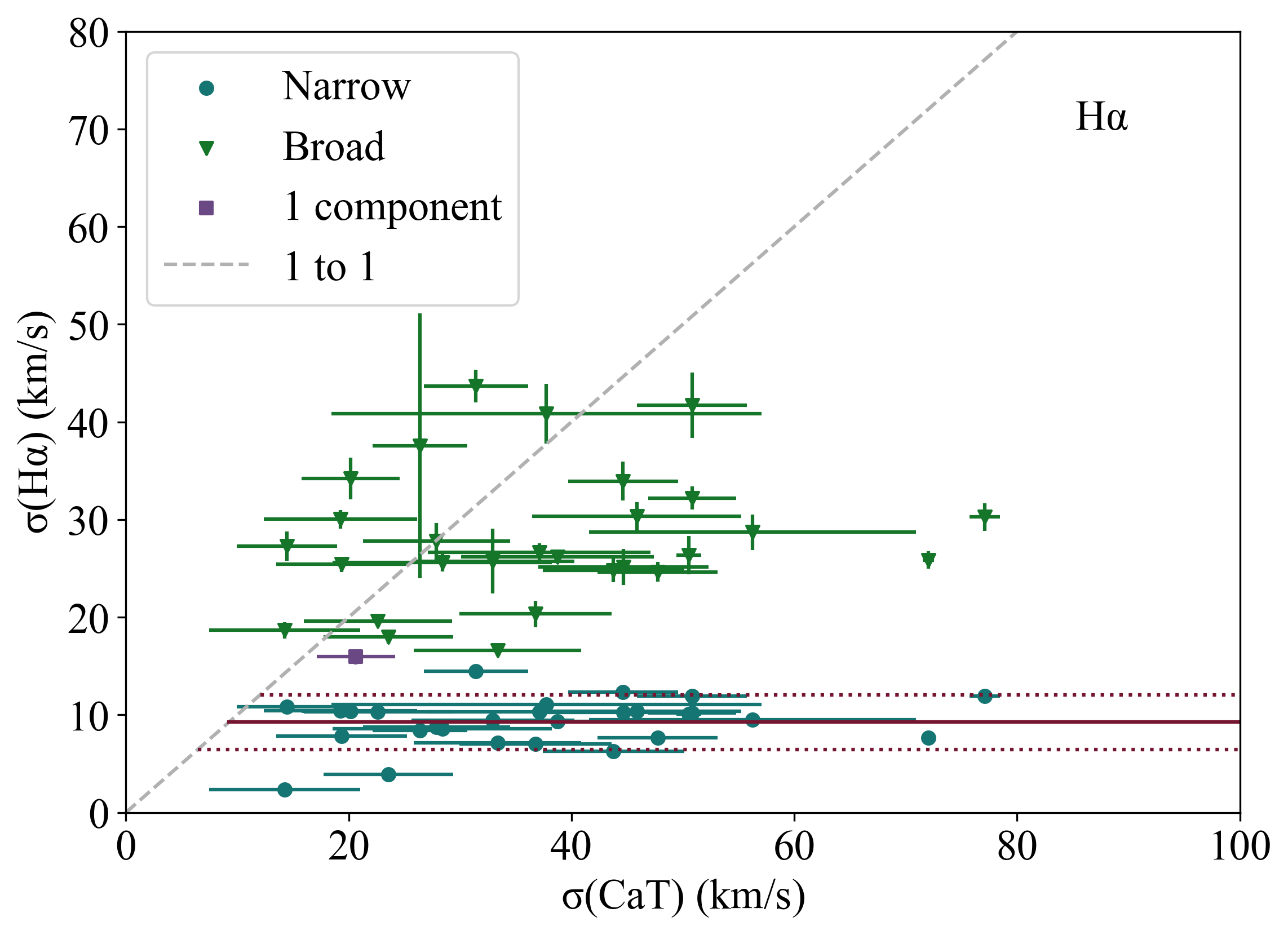}
\includegraphics[width=\columnwidth]{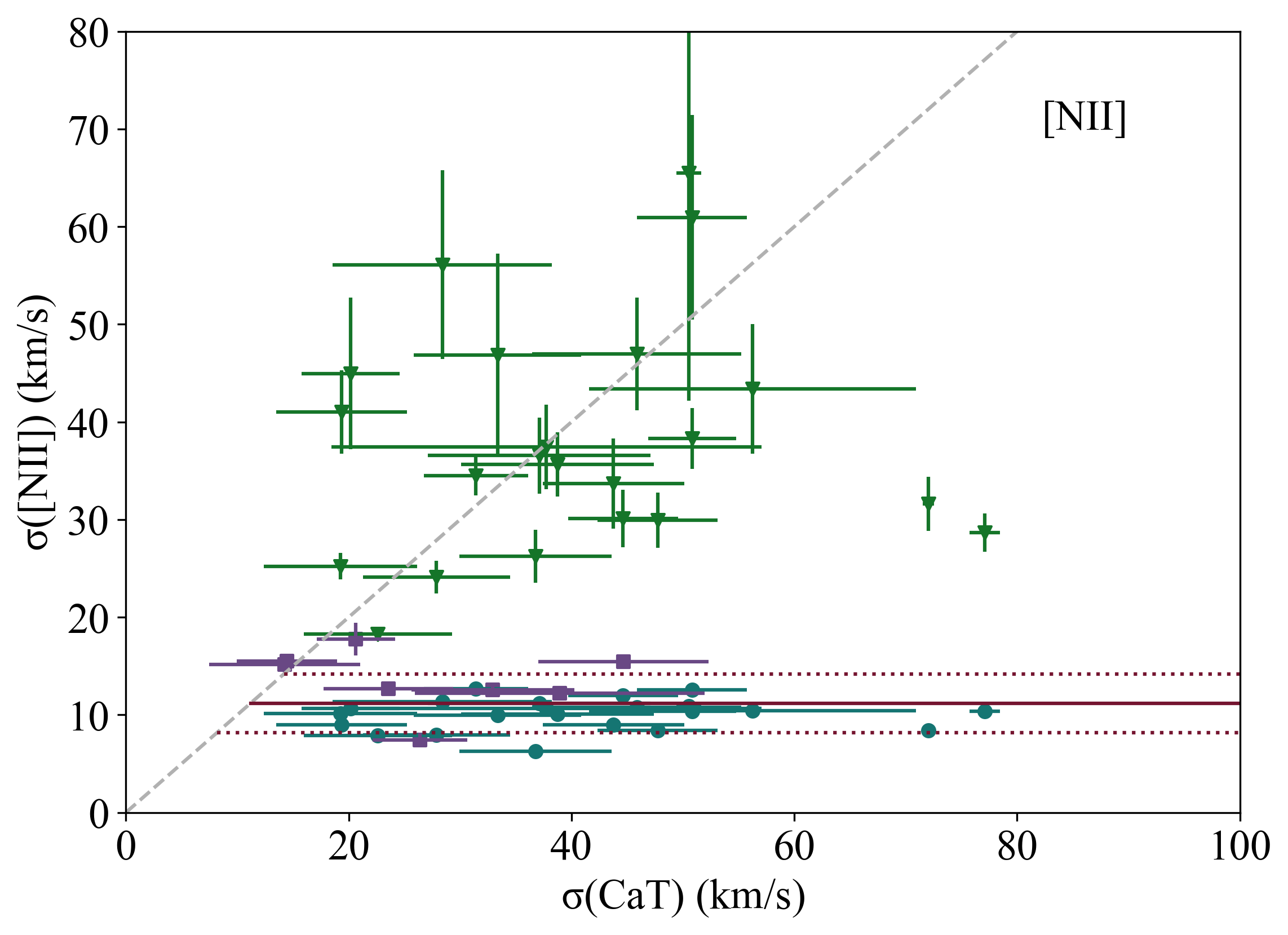}
\includegraphics[width=\columnwidth]{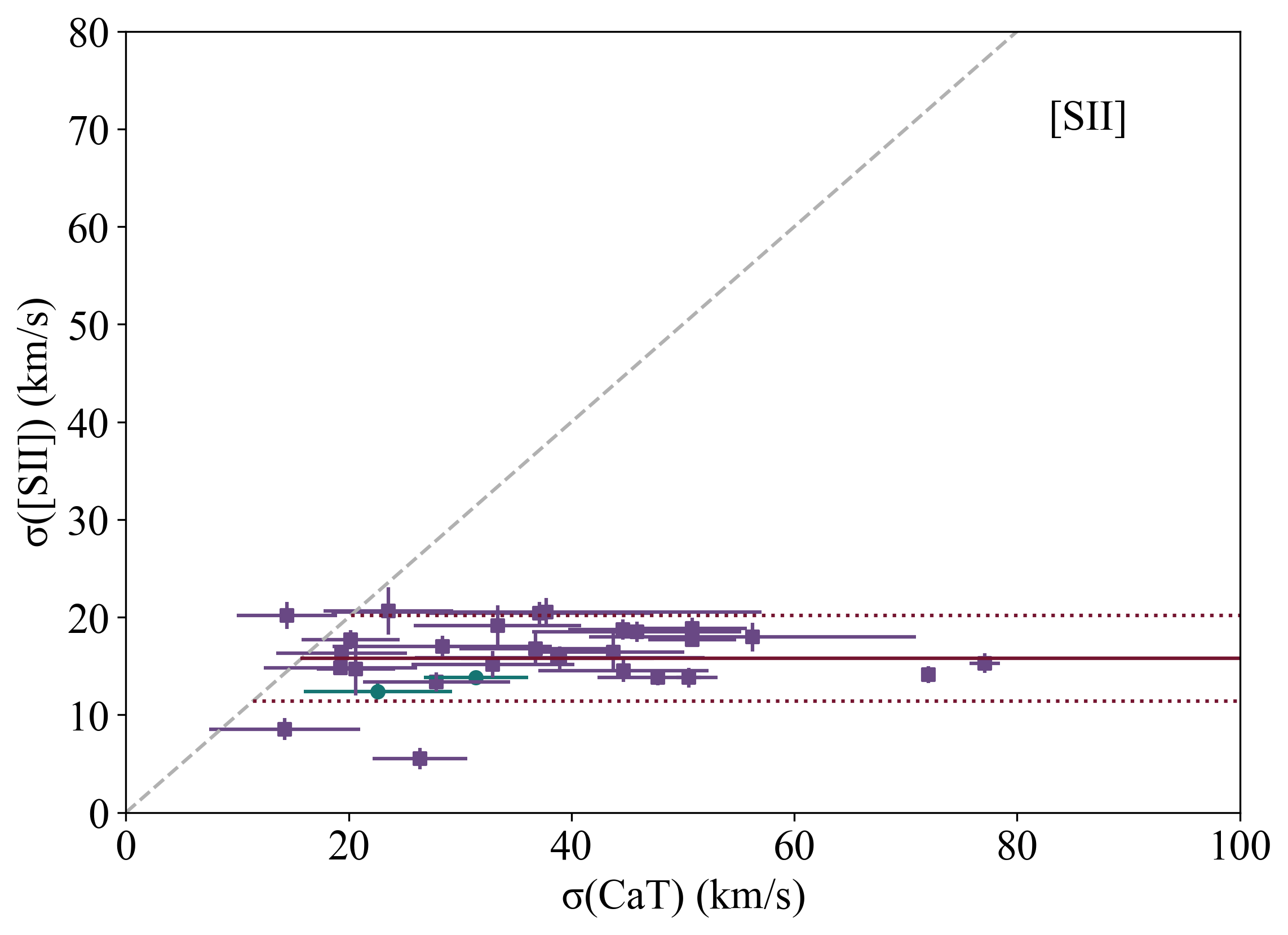}
\includegraphics[width=\columnwidth]{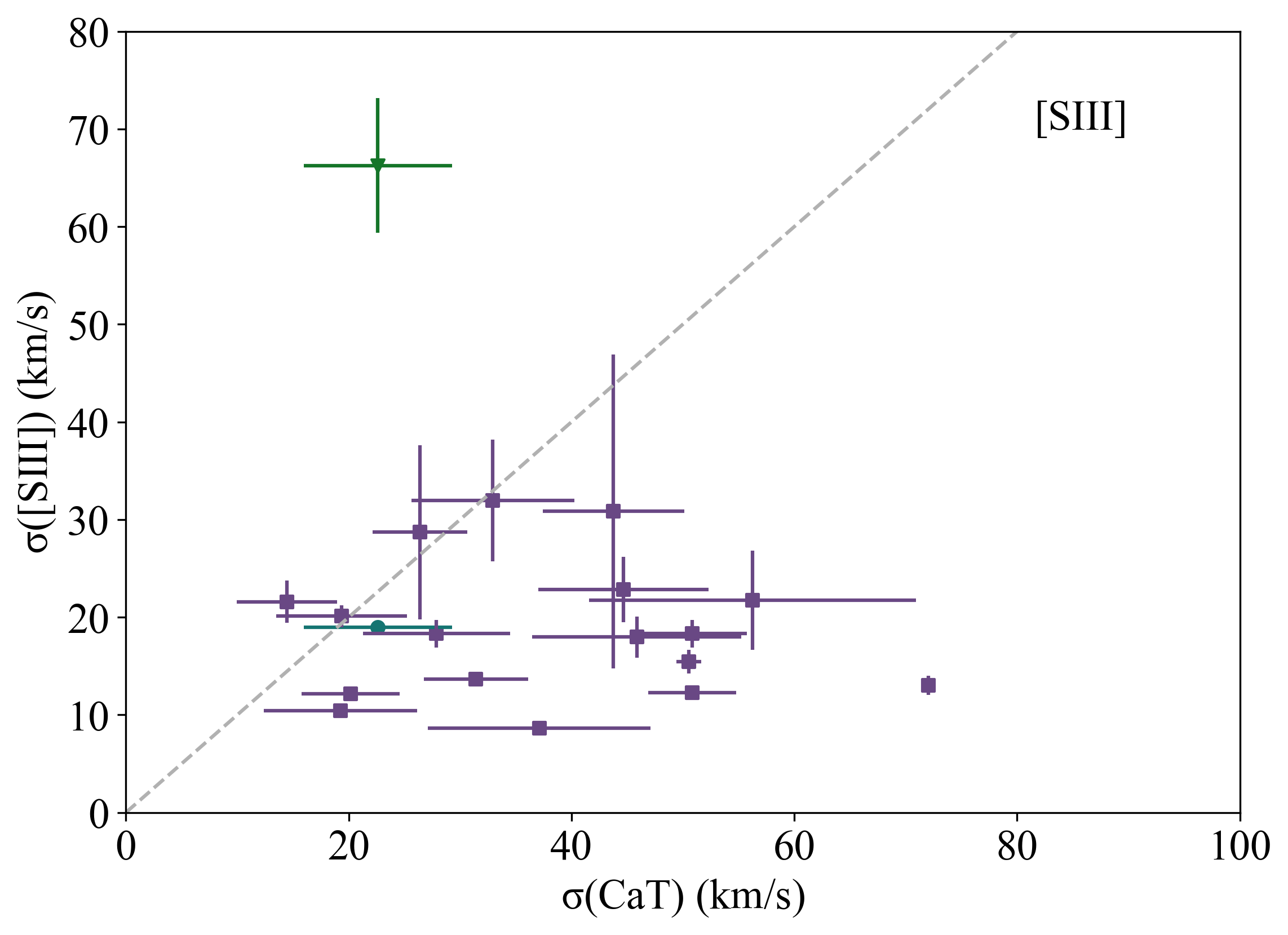}
\caption{Relation between the velocity dispersion of stars (CaT) and ionised gas (H$\alpha$, [NII]$\lambda $ 6584 \AA , [SII]$\lambda $ 6731 \AA , and [SIII]$\lambda$ 9532 \AA\ emission lines) for the selected CNSFRs. The red horizontal solid line represents the average velocity dispersion of the gas narrow component; the dashed red lines correspond to the $\pm$ 1 $\sigma$ value.}
 \label{fig:nebular_lines_sigmas}
\end{figure*}

Figure \ref{fig:nebular_lines_sigmas} shows the comparison between the derived velocity dispersion for stars and gas in the selected CNSFRs. Gas narrow and broad components are represented by blue dots and green triangles, respectively, while regions showing a single component are  represented by purple squares. The dashed grey line shows the one-to-one relation. The top left panel of this figure shows the results for the H$\alpha$ emission line. We can see that the narrow components have velocity dispersions lower than the stellar values and in a narrow range with a mean value of $\overline{\sigma}_{H\alpha}$ = 9.26 km/s, shown by a red solid line; the dotted red lines represent the standard deviation of the distribution. Regions that have only one component show a similar behaviour.  On the other hand, the broad components lie slightly above the one-to-one relation in regions with low star velocity dispersions. This behaviour seems to be lost for higher values of $\sigma_{CaT}$. The top right panel of the figure shows the results for [NII], which show the same behaviour as the H$\alpha$ line. In this case, the mean velocity dispersion of the narrow component is $\overline{\sigma}_{[NII]}$ = 11.2 km/s and the standard deviation is slightly larger. For the broad component the same trend seen for the H$\alpha$ line is easily recognizable. The only component identified in the [SII] lines, shows the same velocity dispersion associated with the gas narrow component, showing a mean value of  $\overline{\sigma}_{[SII]}$ = 15.8 km/s (see the lower right panel of the figure), while for the [SIII] lines the velocity dispersions show a more complex behaviour, as shown in the lower right panel of the figure; the single component present shows a wider velocity dispersion distribution, although mostly below the one-to-one line. The only region with a broad component in [SIII] is R1 appearing in the top left corner of the diagram, with a gas velocity dispersion of 65 km/s (this region is close to supernova SN 1993R).

Similar results to those presented above have also been found in other extragalactic HII region studies in the literature. For example in \citet{2010MNRAS.406.1094F}, where some giant HII regions in the NGC~7479 spiral galaxy were analysed using echelle spectroscopy data with a similar spectral resolution to that of MEGARA. In this work, region III is reported to show a broad Gaussian component in the H$\alpha$ and [NII] emission lines with a velocity dispersion in the range between 37.8 km/s and 51.2 km/s, while the narrow component is around 10 km/s and no broad component was found in the [OIII] profile whose velocity profile can be fitted with a Gaussian with a value of $\sigma$ around 20 km/s. 
 In addition, CNSFRs of different galaxies studied in
\citet{2009MNRAS.396.2295H,2010MNRAS.402.1005H}  show broad components in the emission lines at the same radial velocity as the narrow ones within the errors; the  authors identified the narrow and broad components with ionised gas supported by rotation and dynamical pressure respectively. Interestingly, the CNSFR rings in both galaxies are thought to have been produced in a minor merger event.

A diversity of line broadening mechanisms have been proposed to explain  the line profiles of HII regions. Winds associated with WR stars or young  main sequence O stars \citep[i. e.][]{1987MNRAS.226...19D,1996MNRAS.279.1219T} could produce them.  However, the expected broadening due to WR winds would, in general, be greater than observed in our clusters since these stars can have winds of thousands of km/s. On the other hand, we  estimated the number of O stars that may be present in the CNSFRs of this study. To do so  we  used the \citet[][m$_{low}$ = 0.85M$_\odot$, m$_{up}$ = 120 M$_\odot$]{Salpeter1955} IMF and models from \cite[][Z$_\odot$, log(g) = 4, R$^*$ = 7.1 - 10.6 R$_\odot$]{2002MNRAS.337.1309S}, which derive the following equation:
\begin{equation}
N_O = 6.437\cdot 10^{-5} M_{cluster}.
\end{equation}
\noindent Here M$_{cluster}$ is expressed in solar masses. For the clusters with derived dynamical mass  this number is between 1 and 38, with a mean value of $\sim $ 10 stars. However, normal O stars have winds with velocities $\sim$ 10$^3$ km/s, depending on the density of the environment. Thus, they would also provide a broadening larger than  observed. In conclusion, we rule out stellar winds as the main broadening mechanism in our CNSFRs. 

A second alternative could be related to the interaction of the hot phase of the ISM with cooler gas knots forming a turbulent mixing layer with velocities of about 100 to 
400 km/s \citep{2007MNRAS.381..913W,2007ApJ...671..358W}; Champagne flows arising from the photo-erosion of denser clumps of gas also would produce velocities larger than observed although they can contribute to the turbulence of the ISM \citep{1985A&A...145...70T}. 

Another possible explanation proposed in the literature is the turbulence of the interstellar gas. In 1994, \citet{1994ApJ...425..720C} studied the 30 Doradus giant HII region, in the Large Magellanic Cloud. The outer regions of the nebula have a smooth turbulent velocity field with velocities of 13 to 17 km/s and, at the central regions, the velocity is dominated by multiple expanding structures from 8.5 to 85 km/s (in units of $\sigma$). They corroborated that the integration of all these structures produced a simple broad Gaussian profile with faint extended wings. Additionally, several fast expanding shells were found with velocities from $\sim$ 40 to 130 km/s. 
\citep{1997ApJ...487..163M} studied the giant HII region NGC~604, in M33 finding structure in the velocity field projected along the line of sight that could be attributed to turbulence and suggesting that this turbulence could be powered by recent star formation processes. According to this, the broad component of velocity dispersions found in our particular case look similar to those produced by interstellar gas turbulence.

However, the works just mentioned above lack the information related to the comparison between gas and stars velocity dispersions, and this fact provides additional important information: the stars and the gas associated with the broad component of the emission lines showing similar values of velocity dispersion, and sharing a common kinematics, can be explained if they respond to the same gravitational potential of the cluster.  From the results obtained in the first paper in this series, we concluded that ring clusters are composed of a young non-ionising intermediate age population about 300 Ma old, plus an ionising population of 4.7 Ma in a 4:1 ratio. 
In this scenario, the ring clusters would have formed almost simultaneously in a first star formation episode. The stars in such a population would return part of their gas to the interstellar medium during their evolution. A fraction of this gas could escape from the cluster, while some of it would be unable to do so thus remaining trapped in the cluster \citep{2018MNRAS.478.5112S}. This gas would feel the same gravitational potential as the cluster stars, consistent with our results. A fraction of the gas that cooled down would form a new stellar population, which would be responsible for the ionising photons. In this scenario, the narrow gas component would be associated with the HII region ionisation shell. According to the results presented here  (see Sect. \ref{sec:Dynamical_masses:MEGARA}, this young ionising stellar population accounts for between 0.15 and 7.07 \% of the total dynamical mass of the cluster. 


\section{Conclusions}
In this third paper in the series, we analysed the kinematics of gas and stars in the circumnuclear ring of NGC~7742 estimating the dynamical masses of their CNSFRs from high spectral resolution data. 
We  extracted spectra for each region  using the apertures selected from H$\alpha$ images. 
The analysis of the emission line profiles,  
show the presence of two different kinematical components with the same radial velocity. These two different components were   found in the H$\alpha$ emission line in $\sim$ 90 \% of the analysed regions;  most of them also showed the same component  in the [NII] emission lines. The narrow component has an almost constant value between 10 km/s and 15 km/s, while the broad component velocities are between 20 and 45 km/s.

In spite of the contribution to the nebular continuum, the CaT lines are seen almost undiluted pointing to the stellar population being dominated by red giant or supergiant stars. 

The sizes of the star clusters were  measured from a F555W WFPC2-HST image. About $\sim$ 60 \% show a single star-forming knot;  the rest of them show between two and four knots each. 
Cluster velocity dispersions were derived from the CaT absorption lines using a modified version of the Tonry \& Davis cross-correlation technique. These velocities together with  the measured sizes were   used to calculate the cluster dynamical masses under the assumption that the clusters are virialised. The masses were   found to be between 2.5 $\times$ 10$^6$ and 1.0 $\times$ 10$^8$ M$_\odot$, values similar to those found in the literature for circumnuclear clusters. 

Finally, a comparison of the gas and star velocity dispersion, seldom found in the literature, was  made revealing a complex internal structure of the gas. A scenario was   put forward suggesting that the circumnuclear ring clusters   formed simultaneously in a first star formation episode whose  high-mass stars could have returned part of their gas to the interstellar medium during their evolution. A fraction of this initial ejecta  remained trapped in the  gravitational potential of the cluster. A fraction of the  trapped gas   cooled down and formed the young stellar population responsible for the ionisation of the gas currently observed. This young stellar population accounts for between 0.15 and 7.07 \% of the total dynamical mass of the cluster. 

\begin{acknowledgements}
This research has made use of the services of the ESO Science Archive Facility and NASA’s Astrophysics Data System Abstract Service. It is based on observations collected at the European Organisation for Astronomical Research in the Southern Hemisphere under ESO programme 60.A-9301(A) and data products created thereof. Also we have used observations obtained with the NASA/ESA HST and also from the \textit{Hubble} Legacy Archive, which is a collaboration between the Space Telescope Science Institute (STScI/NASA), the Space Telescope European Coordinating Facility (ST-ECF/ESA), and the Canadian Astronomy Data Centre (CADC/NRC/CSA). 
This work has been supported by Spanish grants from the  former Ministry of Economy, Industry and Competitiveness through the MINECO-FEDER research grants AYA2016-79724-C4-1-P, and PID2019-107408GB-C42 and the present Ministry of Science and Innovation through the research grant PID2022-136598NB-C33 by MCIN/AEI/10.13039/501100011033 and 
by “ERDF A way of making Europe”.
S.Z. acknowledges the support from contract: BES-2017-080509 associated with the first of these grants. This work is based on observations made with the Gran Telescopio Canarias (GTC), installed at the Spanish Observatorio del Roque de los Muchachos of the Instituto de Astrofísica de Canarias, on the island of La Palma. The data was obtained with the instrument MEGARA funded by European Regional Development Funds (ERDF), through Programa Operativo Canarias FEDER 2014–2020.
\end{acknowledgements}

%
\bibliographystyle{aa} 
\bibliography{Article} 
%


\end{document}


\title{Revising the cross correlation technique at high spectral resolution}


   \author{S. Zamora
          \inst{1} \inst{2}\fnmsep\thanks{PhD fellow of Ministerio de Educación y Ciencia, Spain, BES-2017-080509, CEAL-AL/2017-02}
          \and
          A. I. Díaz \inst{1}\inst{2}
          }

   \institute{Departamento de Física Teórica, Universidad Autónoma de Madrid, 28049 Madrid, Spain
         \and
             CIAFF, Universidad Autónoma de Madrid, 28049 Madrid, Spain
             }

   \date{Received XXX; accepted YYY}

 
  \abstract
   {
   Cross-correlation techniques have been used since 1974 and, since 1979, the analysis based on the Fourier Method has been applied. However, we are currently obtaining data with spectral resolution higher than those for which this technique was developed, hence some revision seems timely.  
   }
   {
   The principal aim of this work is to adapt Tonry and Davis' method and implementing it for the treatment of very high spectral resolution data. 
   }
   {
    We have applied this technique to two different sets of spectroscopic data of moderate and high resolutions obtained with the MUSE and MEGARA spectrographs respectively. Using stellar spectra obtained with these two instruments (i) we have optimised the input parameters; (ii) we have analysed the method assumptions; and (iii) we have compared the results for the two sets of data. 
   }
   {
   The optimal method parameters applied to MUSE data are $k_{min}$ $\sim$ 3, $k_{max}$ $\sim$ 60 and 512 bins, which correspond to a uniform velocity shift value of $\Delta $v = 27.1 km/s. For MEGARA data, we propose the values $k_{min}$ $\sim$ 3, $k_{max}$ $\sim$ 350 and 4096 bins finding that the cross-correlation function lost its Gaussian behavior at higher resolutions. Thus, we have developed an equivalent mathematical method that can be used for this kind of data. Additionally, the velocity dispersion error analysis suggests that the greatest error introduced in this method is due to the subtraction or masking of the nebular lines.
   }
   {
   For the application cross-correlation techniques to high spectral resolution data, we propose to calculate the galaxy-galaxy and star-galaxy correlations, with widths $\mu_{gg}$ and $\mu_{gt}$ respectively. Then, the width of the broadening function can be calculated as $\sigma = \sqrt{\mu_{gg}^2 - \mu_{gt}^2}$.
   }

\keywords{galaxies: star clusters: general, galaxies: starburst, techniques: imaging spectroscopy, Astronomical instrumentation, methods, and techniques}

   \maketitle
%

\section{Introduction}
Cross-correlation techniques have been applied since several decades ago \citet{1974A&A....31..129S} in order to derive objects’ redshifts and velocity dispersions by correlating the object's spectrum with that of a reference star and measuring respectively the peak center and width of the found cross-correlation function. In \citet{1979AJ.....84.1511T} a detailed correlation analysis based on the Fourier Method was presented. The use of the  Fourier Transform algorithm reduces the operational and computing times, providing a straightforward determination of the cross correlation function. Additionally, an internal error measure of the correlation peak can be directly performed. 

This technique can be applied to a large number of astronomical problems. Examples of these are: the derivation of the stellar velocity rotation, the gas and star  velocity dispersion in different types of galaxies (including broad and narrow line active galaxies), or the study of regions with active star formation  \citep[i.e.][]{2007MNRAS.378..163H,1995ApJS...99...67N}.

Over at least half-a-century, numerous thorough method reviews have been presented by various authors, beginning with \citet{1978ApJ...221....1D} and followed by \citet{2001BSAO...51...11M} or \citet{2006ApJ...641..117G}. However, we are currently obtaining data with characteristics very different  from those used up to now that include a good number of spectrographs of very high spectral resolution. At this high resolutions, the cross correlation technique assumptions may not be correct, hence a revision of the method is required. 


In this work, we have tested the widely used Tonry and Davis' technique \citep{1979AJ.....84.1511T} using data of moderate and high spectral resolutions obtained with two spectrographs currently in use: MUSE (R ~ 2500) and MEGARA (R ~6000 - 20000) adapting this method to the analysis of high spectral resolution data. We have chosen these two instruments just for illustration purposes although the results obtained can be generalised to any other since they do not depend on a given spectrograph but only on the spectral resolution.  

In Section 2, we describe the observations used and the spectral resolution expected for each set of the data used. The cross correlation technique is introduced in Section 3 including the necessary mathematical concepts, the presentation of the original method, and the developing of the one proposed in this work, as well as the steps followed in the spectrum preparation required prior to the  application of the method. Our analyses and results are presented in Section 4 for the MUSE and MEGARA data respectively. A discussion of the method assumptions at different spectral resolutions, the comparison between the traditional methodology and the new one presented here, and the analysis of the different errors involved in them are given in Section 5. Finally, the summary and conclusions of this work are given in Section 6.

\section{Observations}
\label{sec:observations}
In this work, we have used data from two different instruments in order to compare the traditional Tonry and Davis' method \citep{1979AJ.....84.1511T} and the proposed one at high spectral resolutions. We have used data from the Multi-Unit Spectroscopic Explorer \citep[MUSE][]{MUSE} and from the Multi-Espectrógrafo en GTC de Alta Resolución para Astronomía \citep[MEGARA][]{2018SPIE10702E..16C}. Both instruments are integral-field spectrographs (IFS) and are attached to one of the Very Large Telescopes (VLT) of the European Southern Observatory (ESO, Chile) and at the 10.4-m GTC telescope at La Palma Observatory (Spain) respectively. MUSE has a nominal dispersion of 1.25 \AA /pixel with a spectral resolving power, R$_{FWHM}$ = $\lambda / \Delta \lambda$, from 1770 (at 4800 \AA ) to 3590 (at 9300 \AA ) in the blue and red arms respectively. It covers the visible wavelength range from 4800 \AA\ to 9300 \AA\ . On the other hand, MEGARA has three observing modes: low resolution LR, medium resolution MR and high resolution HR providing R values from $\sim$ 6000, 12000 and 20000 respectively, covering from 3654 \AA\ to 9635 \AA\ in different dispersion elements.

\input{Tables/MUSE_stars}
\begin{table*}
\centering
\caption{Selected giants and super-giants MEGARA stars from \citet{2020MNRAS.493..871G}.}
\label{tab:MEGARA_stars}
\begin{tabular}{ccccccc}
\hline
Name     & RA J2000 (deg)         & DEC J2000 (deg)        & Spectral type     & Effective Temperature (K) & log(g) & {[}Fe/H{]} \\
\hline
HD042983 & 06:13:52.8 & +02:48:31.9 & K0          & 4633 & 3.23 & 0.00       \\
HD105087 & 12:06:00.9 & +14:38:56.7 & K0          & 4966 & 4.16 & -0.43      \\
HD215704 & 22:46:20.4 & +50:12:35.9 & K0          & 5418 & 4.20 & 0.07       \\
HD040801 & 06:03:17.9 & +42:54:41.5 & K01II       & 4740 & 2.94 & -0.04      \\
HD045829 & 06:30:02.3 & +07:55:16.0 & K0Iab       & 4500 & 0.20 & 0.00       \\
HD027371 & 04:19:47.6 & +15:37:39.5 & K0III       & 4956 & 2.71 & 0.07       \\
HD045410 & 06:30:47.1 & +58:09:45.5 & K0III       & 4838 & 2.34 & 0.17       \\
HD100696 & 11:36:02.8 & +69:19:22.6 & K0III       & 4890 & 2.27 & -0.25      \\
HD107328 & 12:20:21.0 & +03:18:45.3 & K0III       & 4380 & 2.39 & -0.48      \\
HD131111 & 14:50:29.6 & +37:16:19.4 & K0III       & 4710 & 3.11 & -0.29      \\
HD147677 & 16:22:05.8 & +30:53:31.2 & K0III       & 4910 & 2.98 & -0.08      \\
HD063302 & 07:47:38.5 & -15:59:26.5 & K2Iab       & 4500 & 0.20 & 0.12       \\
HD055280 & 07:15:54.9 & +59:38:14.9 & K2III       & 4623 & 2.54 & 0.08       \\
HD072184 & 08:32:55.0 & +38:00:58.9 & K2III       & 4624 & 2.61 & 0.12       \\
HD115136 & 13:13:28.0 & +67:17:16.6 & K2III       & 4541 & 2.40 & 0.05       \\
HD017506 & 02:50:41.8 & +55:53:43.8 & K3I         & 3500 & 1.00 & 0.09       \\
HD083425 & 09:38:27.3 & +04:38:57.4 & K3III       & 4120 & 1.77 & -0.29      \\
HD125560 & 14:19:45.2 & +16:18:25.0 & K3III       & 4426 & 2.42 & 0.00       \\
HD034255 & 05:20:22.6 & +62:39:13.4 & K4I         & 3000 &      & -0.23      \\
HD126271 & 14:24:18.3 & +08:05:04.6 & K4III       &      &      & -0.12      \\
HD131507 & 14:51:26.4 & +59:17:38.4 & K4III       & 4140 & 1.99 & -0.20      \\
HD149161 & 16:32:36.3 & +11:29:16.9 & K4III       & 3910 & 1.39 & -0.17      \\
HD044274 & 06:21:02.9 & +02:34:07.6 & M           & 4122 & 0.50 & 0.00       \\
HD060501 & 07:35:03.0 & +06:11:39.7 & M           & 4003 & 0.50 & 0.00       \\
HD042543 & 06:12:19.1 & +22:54:30.7 & M0Iab       & 3615 & 0.00 & -0.42      \\
HD039801 & 05:55:10.3 & +07:24:25.4 & M1-M2Ia-Iab & 3547 &      & 0.03       \\
HD035601 & 05:27:10.2 & +29:55:15.8 & M1.5Ia      & 3700 & 0.20 &            \\
HD044537 & 06:24:53.9 & +49:17:16.4 & M2I         & 3055 &      & 0.08       \\
HD078712 & 09:10:38.8 & +30:57:47.3 & M6III       & 3210 & 0.00 & -0.11      \\
HD196610 & 20:37:54.7 & +18:16:06.9 & M6III       &      &      &           \\
\hline
\end{tabular}
\end{table*}

In this work, we have used late-type giant and supergiant stars as templates. The sample stars have been selected from the MUSE and MEGARA stellar libraries presented in \citet{2019A&A...629A.100I} and \citet{2020MNRAS.493..871G} respectively. Tables \ref{tab:MUSE_stars} and \ref{tab:MEGARA_stars} show the characteristics of each of the selected stars listing in columns 1 to 7: (1) the star name; (2,3) the right ascension and declination of the star; (4) its spectral type; (5) its effective temperature; (6) its surface gravity in logarithmic units; and (7) its metallicity.

The Ca II $\lambda \lambda $ 8498, 8542, 8662 \AA\ triplet lines (CaT) have often been used to measure the velocity dispersion of stars in galaxies. The use of these features with respect to other ones present in galaxy spectra presents some advantages. First, the velocity resolution at longer wavelengths is higher than at the bluer part of the spectrum for the same spectral dispersion. At these wavelengths, we expect $\sim $ 37.7 km/s and $\sim$ 6.2 km/s spectral resolutions in $\sigma$ for the MUSE and MEGARA (using the HR-I mode, VPH 863) data respectively. Second, the measurement of these lines is not affected by the presence of TiO bands. And third, the CaT lines are dominated by young supergiant stars, much more luminous than red giants, whose strength increases with decreasing surface gravity \citep{1989MNRAS.239..325D} (although they depend on metal abundance in the low metallicity regime, at moderate to high metallicities surface gravity becomes the dominant parameter governing these strengths). Also, they are less affected by the possible presence of an AGN since the dilution of stellar features due to the presence of a power-law continuum is lower at red wavelengths. Thus, they are especially suitable for the study the velocity dispersions of young star clusters located at the central regions of galaxies, as circumnuclear star-forming regions (CNSFRs).

\begin{figure}
\centering
\includegraphics[width=\columnwidth]{Images/Plot_stars_Stars_MUSE.png}
\caption{MUSE Giant and supergiant stars (upper and lower spectra respectively, see text).}
\label{fig:MUSE_stars}
\end{figure}

\begin{figure}
\centering
\includegraphics[width=\columnwidth]{Images/Plot_stars_Stars_MEGARA_giant.png}
\includegraphics[width=\columnwidth]{Images/Plot_stars_Stars_MEGARA_supergiant.png}
\caption{MEGARA giant and supergiant stars (upper and lower panels respectively, see text).}
\label{fig:MEGARA_stars}
\end{figure}

Figures \ref{fig:MUSE_stars} and \ref{fig:MEGARA_stars} show the stellar spectral range used in this work. Red and blue spectra correspond to red giant and supergiant stars respectively and at the end of each series the mean spectrum which has been used as template is shown. All the spectra have been corrected for velocity shifts so that no artificial broadening was introduced in the final templates. For our analysis, we have used a mean stellar template computed after aligning all selected spectra and verifying that no apparent broadening is introduced in the procedure. We have checked that the use of different stars of the same spectral type does not introduce errors in the correlation results. Nevertheless, the use of only two stellar types as templates can introduce errors in the velocity dispersion measurements of a given star cluster or galaxy due to the presence of different stellar types with different luminosities. The possible mismatch between the stellar template and the cluster or galaxy lines has been minimised by using the CaT lines since they are very strong in most stars except for the hottest ones. However, it is important to emphasise that our work constitutes only a method revision and its application should be adapted to each particular scientific case.

\section{Cross Correlation Technique}
\subsection{Correlation theoretical concepts} \label{sec:teoria}
The convolution product has been applied with discrete Fourier transforms, using the Convolution, the Correlation and the Parseveral Theorems. For two generic functions, s(t) p(t), these theorems can respectively be abbreviated as: 
\begin{displaymath}
F\left[s(t)\ast p(t)\right](k) = S(k)\cdot P(k)
\end{displaymath}
\begin{displaymath}
F\left[s(t)\otimes p(t)\right](k)=P(k)\cdot S^*(k)
\end{displaymath}
\begin{displaymath}
\displaystyle\sum _{n} s(n)^2 = \frac{1}{N}\displaystyle\sum _{k} |S(k)|^2
\end{displaymath}
\noindent where $F$ denotes the discrete Fourier transform, $\ast$ means the convolution product, $\otimes$ means the cross correlation product, $S(K)$ and $P(k)$ are the Discrete Fourier Transforms of $s(t)$ and $p(t)$ functions respectively, and $S*(K)$ and $|S(k)|$ are the complex conjugate and the module of $S(k)$ respectively. The spectra will be considered periodic in N in order to calculate the discrete Fourier transforms.

The method is based on the idea that galaxy spectra can be represented by a star spectrum convolved with a broadening function: 
\begin{equation}\label{eq:gn}
g(n)\sim \alpha [t(n)\ast b(n-\delta)]
\end{equation}
\noindent where $\ast$ stands for the convolution product, $g(n)$ is the galaxy spectrum, $\alpha $ is the number of stars, $t(n)$ is the template spectrum, $b(n)$ is the broadening function and $\delta$ is the offset of the broadening function with respect to the template. This equation can be understood as the galaxy spectrum being a sum of different star spectra with different velocity offsets.

The broadening function will be assumed to be a Gaussian:
\begin{equation}
b(n) = \frac{1}{\sqrt{2\pi } \sigma} e^{-\frac{n^2}{2\sigma ^2}}
\end{equation}
\noindent whose Fourier transform is: 
\begin{displaymath}
B(k) = \displaystyle\sum _{n} \frac{1}{\sqrt{2\pi } \sigma} e^{-\frac{n^2}{2\sigma ^2}} e^{-\frac{2\pi i n k}{N}} = \frac{e^{-\frac{\left(2 \sigma \pi k\right)^2}{2N^2}}}{\sqrt{2\pi } \sigma}  \displaystyle\sum _{n}  e^{-\frac{\left[n + \left(\frac{2 \sigma^2 \pi i k}{N}\right)\right]^2}{(\sqrt{2}\sigma )^2}}
\end{displaymath}
Taking into account the approximation $\sum e^{n^2/\sigma ^2}\sim \sqrt{\pi} \sigma$ this can be writen as:
\begin{equation}\label{B(k)}
B(k) \sim  e^{-\frac{\left(2 \sigma \pi k\right)^2}{2N^2}}
\end{equation}
Also, a Gaussian amplitude of stellar Fourier transform can be assumed:
\begin{equation}\label{T(k)}
|T(k)| = \sigma _t \left(\frac{2\pi N \tau}{\sqrt{\pi}}\right)^{1/2}e^{-\frac{(2\pi \tau k)^2}{2N^2}}
\end{equation}


\subsection{Cross correlations methods}\label{method}
In this work, we have calculated three cross correlations: (i) the correlation of the galaxy spectrum with itself; (ii) the correlation of the star spectrum with itself; and (iii) the correlation of the galaxy spectrum with the star spectrum. We have used the following normalisation:
\begin{displaymath}
\sigma_g^2 = \frac{1}{N} \displaystyle\sum _{n}g(n)^2 \qquad \sigma_t^2 = \frac{1}{N} \displaystyle\sum _{n}t(n)^2
\end{displaymath}

The correlation of the galaxy spectrum with itself can be calculated as:
\begin{displaymath}
c_{gg}(n) = g(n)\otimes g(n)|_{norm} = \frac{1}{N  \sigma_g^2}\displaystyle\sum _{m} g(m)\cdot g(m-n) 
\end{displaymath}
In the Fourier form:
\begin{equation}\label{eq:c1}
\begin{split}
C_{gg}(k) = F[c_{gg}(n)] = F[g(n)\otimes g(n)]_{norm} 
= \frac{1}{N \sigma_g^2} G(k)\cdot G(k)* \\
=\frac{1}{N \sigma_g^2} |G(k)|^2= \frac{\alpha ^2}{N\sigma_g^2} |T(k)\cdot B(k)|^2
=\frac{\alpha ^2}{N\sigma_g^2} |T(k)|^2\cdot |B(k)|^2\\= \frac{2\pi \tau \alpha^2 \sigma_t^2}{ \sigma_g^2\sqrt{\pi}}\cdot exp\left[{\frac{-k^2}{2 \cdot \left(\sfrac{N}{2\pi\sqrt{2\tau^2 +2\sigma^2}}\right)^2}}\right]
 \end{split}
\end{equation}

The correlation of the star spectrum with itself can be calculated as:
\begin{displaymath}
c_{tt}(n) = t(n)\otimes t(n)|_{norm} = \frac{1}{N  \sigma_t^2}\displaystyle\sum _{m} t(m)\cdot t(m-n) 
\end{displaymath}
In the Fourier form:
\begin{equation}\label{eq:c2}
\begin{split}
C_{tt}(k) = F[c_{tt}(n)] = F[t(n)\otimes t(n)]_{norm} 
= \frac{1}{N \sigma_t^2} T(k)\cdot T(k)*\\
=\frac{1}{N \sigma_g^2} |T(k)|^2=  \frac{2\pi \tau}{ \sqrt{\pi}}\cdot exp\left[\frac{-k^2}{2 \cdot \left(\sfrac{N}{2\pi\sqrt{2}\tau}\right)}\right]
 \end{split}
\end{equation}

Finally, the correlation of the galaxy spectrum with the star spectrum can be calculated as:
\begin{displaymath}
c_{gt}(n) = g(n)\otimes t(n)|_{norm} = \frac{1}{N\sigma_t \sigma_g}\displaystyle\sum _{m} g(m)\cdot t(m-n) 
\end{displaymath}
In the Fourier form:
\begin{equation}\label{eq:c3}
\begin{split}
C_{gt}(k) = F[c_{gt}(n)] = F[g(n)\otimes t(n)]_{norm} 
= \frac{1}{N\sigma_t \sigma_g} G(k) \cdot T^*(k) \\= \frac{1}{N\sigma_t \sigma_g} F[\alpha [t(n)\ast b(n-\delta)]] \cdot T^*(k) 
=\frac{\alpha}{N\sigma_t \sigma_g} B(k)\cdot |T(k)|^2 \\= \frac{2\pi \tau \alpha \sigma_t}{ \sigma_g\sqrt{\pi}}\cdot exp\left[{\frac{-k^2}{2 \cdot \left(\sfrac{N}{2\pi\sqrt{2\tau^2+\sigma ^2}}\right)^2}}\right]
 \end{split}
\end{equation}

\subsection{Velocity dispersion from the correlation peak}
A Gaussian behaviour of the correlation peak has been assumed, centered at $\delta$ and with dispersion $\mu$:
\begin{equation}\label{eq:c}
c(n)\sim c(\delta)\cdot e^{-\frac{(n-\delta)^2}{2\mu^2}}
\end{equation}
whose Fourier transform is: 
\begin{displaymath}
C(k) = c(\delta)\displaystyle\sum _{n} 
    e^{-\frac{\delta^2}{2\mu^2}}
    e^{-\frac{\left(n+\frac{2 \mu^2 \pi i k}{N}-\delta \right)^2}{2\mu^2}}
    e^{\frac{\left(\frac{2 \mu^2 \pi i k}{N}-\delta \right)^2}{2\mu^2}}
\end{displaymath}
Taking into account the approximation $\sum e^{n^2/\sigma ^2}\sim \sqrt{\pi} \sigma$ it can be writen as:
\begin{equation}\label{eq:C(k)}
C(k) \sim  c(\delta) \sqrt{2\pi} \mu_i \cdot e^{-\frac{2  \pi i k \delta_i}{N}} \cdot exp\left[-\frac{k^2}{2\left(\sfrac{N}{2\pi \mu } \right)^2}\right] 
\end{equation}

The comparison between this last equation and the results of the three calculated cross-correlation functions (see Sec. \ref{method}) establishes relationships between the correlation peaks, $\mu _{gg,tt,gt}$, the template width in the Fourier space, $\tau$, and the broadening function width,  $\sigma$. From Eqs. \ref{eq:c1} and \ref{eq:C(k)}:
\begin{equation}\label{eq:mu1}
\mu _{gg}^2 = 2\tau^2+2\sigma^2
\end{equation}
From Eqs. \ref{eq:c2} and \ref{eq:C(k)}:
\begin{equation}\label{eq:mu2}
\mu _{tt}^2 = 2\tau^2
\end{equation}
From Eqs. \ref{eq:c3} and \ref{eq:C(k)}:
\begin{equation}\label{eq:mu3}
\mu _{gt}^2 = 2\tau^2+\sigma^2
\end{equation}

Combining these three last equations, we can fit a Gaussian to the correlation functions and derive the  broadening function width, which is related to the velocity dispersion to be measured. From Eqs. \ref{eq:mu1} and \ref{eq:mu2}:
\begin{equation}\label{eq:sigma1}
\sigma^2 = \frac{\mu^2_{gg}-\mu^2_{tt}}{2}
\end{equation}
From Eqs. \ref{eq:mu1} and \ref{eq:mu3}:
\begin{equation}\label{eq:sigma2}
\sigma^2 = \mu^2_{gg}-\mu^2_{gt}
\end{equation}
From Eqs. \ref{eq:mu2} and \ref{eq:mu3}:
\begin{equation}\label{eq:sigma3}
\sigma^2 = \mu^2_{gt}-\mu^{2}_{tt}
\end{equation}
This last equation is the one proposed by \citet{1979AJ.....84.1511T}. The other two equations are equivalent to this one being their assumptions and mathematical development analogous to the original method.

\subsection{Spectrum preparation for the correlation analysis}
Each spectrum analysed with the cross-correlation technique should be binned into logarithmic wavelengths in order to get a uniform velocity shift. Let $\lambda$ be the wavelength of the spectrum and n the equivalent bin number; then we can apply the following equation: 
\begin{equation}\label{eq:1}
n(\lambda) = A\cdot ln(\lambda)+B
\end{equation}
where A and B are constants for each spectrum. 

The number of the sampled bins depends on the velocity resolution, $\Delta v$, which we need to obtain. If the wavelength range of the analysed spectrum is between $\lambda _1$ and $\lambda _2$, the minimum number of bins that should be used to not lose spectral resolution can be calculated as: 
\begin{equation}\label{eq:N}
N = \frac{ln(\lambda_ 2/\lambda_ 1)}{ln(\Delta v/c +1)}
\end{equation}
where c is the speed of light in the same units of $\Delta v$. Using this value, we can calculate the A and B constant values from Eq. \ref{eq:1}:
\begin{equation}\label{eq:B}
B = N\cdot \left(1-\frac{ln(\lambda_ 2)}{ln(\lambda_ 1)}\right)
\end{equation}
\begin{equation}\label{eq:A}
A = -\frac{N}{ln(\lambda _1)}\cdot \left(1-\frac{ln(\lambda_ 2)}{ln(\lambda_ 1)}\right)
\end{equation}

Any emission line present in the spectrum should be removed. Also, the continuum spectrum should be subtracted. A second order polynomial is sufficient to model this component since any mistake made will be corrected in the following steps.

The final step in the spectrum preparation is filtering the high and low frequency variations that can affect the velocity dispersion determination. They are associated with noise components and errors in the continuum subtraction respectively. This filtering is performed by applying a band-pass to the Fourier spectrum transform with  minimum and maximum wave numbers, k$_{min}$ and k$_{max}$ which can be calculated as follows:
\begin{equation}\label{eq:kmin}
k_{min} = \frac{N}{2\pi A \cdot ln\left(1-\Delta \lambda^{max}/\lambda _1\right) }
\end{equation}
\begin{equation}\label{eq:kmax}
k_{max} = \frac{N}{2\pi A \cdot ln\left(1-\Delta \lambda^{min}/\left(2 \cdot \lambda _2 \sqrt{2\cdot ln(2)}\right) \right)}
\end{equation}
\noindent where $\Delta \lambda ^{min}$ and $\Delta \lambda ^{max}$ are the minimum and maximum detail that we expect in each spectrum. We have assumed as the minimum value the nominal spectral resolution in \AA , and as the maximum value 10 \AA, a size significantly larger than the width of the absorption lines in each spectrum.

\section{Analysis and results}

We have used the mean spectrum of all stars presented above to calculate the broadening function of MUSE and MEGARA data aiming to check: (i) the traditional methodology proposed by Tonry and Davis \citep[][from Eq. \ref{eq:sigma3}]{1979AJ.....84.1511T} and (ii) the other two proposed in this work (from Eq. \ref{eq:sigma1} and Eq. \ref{eq:sigma2}). Additionally, we have analysed the results of the frequency filtering in the data and we have quantified the effects of the spectrum preparation steps in the results.

\subsection{Application to MUSE data}

First, we have selected wavelengths between 8450\AA\ and 8850 \AA\ in order to exclude the [OI]$\lambda $ 8446 \AA\ and Pa11 lines. We have binned the template spectrum into logarithmic wavelengths using N = 512 bins which corresponds to a uniform velocity shift of $\Delta $v = 27.1 km/s, lower than the nominal velocity resolution of MUSE. Using Eqs. \ref{eq:A} and \ref{eq:B} we have evaluated the constants as A = 11070.03 and B = -100094.31.

Next, we have calculated and subtracted the continuum spectrum, also masking emission nebular lines and absorption stellar lines using widths of 6 \AA\ and 20 \AA\ respectively, and we have fitted a quadratic equation. 

Later, we have used the N value and the selected wavelengths to calculate the band-pass filter constants using Eqs. \ref{eq:kmin} and \ref{eq:kmax}. The selected wave numbers in the Fourier space for MUSE data are $k_{min}$ $\sim$ 3 and $k_{max}$ $\sim$ 60 which correspond to a bin number shift between 1.4 and 26.2. Thus, the Fourier transform of each spectrum  has been multiplied by the following function: 
\begin{equation}
    f(k) =
\begin{cases*}
    0 & if k < 1.5\\\
    k/1.5-1 & if 1.5< k < 3\\\
    1 & if 3< k < 60\\\
    -k/60+2 & if 60< k < 120\\\
    0 & if k > 120
\end{cases*}
\end{equation}

\begin{figure}
\includegraphics[width=\columnwidth]{Images/Filtered_MUSE.png}
\caption{MUSE star spectrum after resampling into logarithmic wavelengths and applying the pass-band filter for the higher frequencies.}
\label{fig:filter_frec_MUSE}
\end{figure}

Figure \ref{fig:filter_frec_MUSE} shows the result of this filtering at high frequencies. The original spectrum and the filtered one are shown as dotted and solid lines respectively. We can see that the filter application removes the noise component of the analysed spectrum.

\begin{figure}
\includegraphics[width=\columnwidth]{Images/Tonry-1979_Fig6_MUSE.png}
\includegraphics[width=\columnwidth]{Images/Tonry-1979_Fig3_MUSE.png}
\caption{Upper panel: Amplitude of the Fourier transform of the MUSE stellar template (red triangles), this template convolved with a Gaussian function (blue triangles) and the cross correlation of the two (purple dots). Lower panel: Fourier transform phase of the same functions with the same colour coding.  }
\label{fig:amplitude_phase_MUSE}
\end{figure}

In order to verify if the frequency filter is well applied, we have simulated a cluster spectrum convolving the stellar template with a known Gaussian function of width $\sigma$ = 1.09 \AA, which corresponds to the MUSE spectral resolution. Figure \ref{fig:amplitude_phase_MUSE} shows, in the upper and lower panels respectively, the phase and the amplitude of the cross-correlation function in the Fourier space as a function of wave number. Red triangles show the template results; blue triangles show the results of the simulated cluster spectrum; and purple dots show the results of the correlation between the two. Vertical grey lines mark the  k$_{max}$ and 2$\cdot$k$_{max}$ values used for the band-pass filter. We can see that the phase becomes random at higher values of the 2$\cdot$k$_{max}$ value while the amplitude is compatible with zero in the same range. Hence, we can conclude that the chosen k$_{max}$ values are correct.

\subsection{Application to MEGARA data}
\label{sec:megara_apli}

In this case, we have selected the same wavelength range although the needed number of bins is now larger, N = 4096 ($\Delta $v = 3.4 km/s), corresponding to a value lower than the velocity resolution of HR-I MEGARA setup. The appropriate binning constants are A = 88560.21 and B = -800754.52. As in the case of MUSE data, nebular and stellar lines have been removed.

Then, we have applied the band-pass filter with wave numbers in Fourier space $k_{min}$ $\sim$ 3 and $k_{max}$ $\sim$ 350 corresponding to bin number shifts between 1.9 and 209.4: 
\begin{equation}
    f(k) =
\begin{cases*}
    0 & if k < 1.5\\\
    k/1.5-1 & if 1.5< k < 3\\\
    1 & if 3< k < 350\\\
    -k/350+2 & if 350< k < 700\\\
    0 & if k > 700
\end{cases*}
\end{equation}

\begin{figure}
\includegraphics[width=\columnwidth]{Images/Filtered_MEGARA.png}
\includegraphics[width=\columnwidth]{Images/Filtered_2_MEGARA.png}
\caption{MEGARA star spectrum after resampling into logarithmic wavelengths and applying the pass-band filter for the higher frequencies (upper panel) and the lower ones (bottom panel).}
\label{fig:filter_frec_MEGARA}
\end{figure}

Figure \ref{fig:filter_frec_MEGARA} shows the result of this filtering on the data. The upper panel shows the effect of the high frequency variation whose behaviour is similar to the case of MUSE data. The lower panel shows the effect of the low frequency filter which corrects the errors in the continuum subtraction as expected.

\begin{figure}
\includegraphics[width=\columnwidth]{Images/Tonry-1979_Fig6_MEGARA.png}
\includegraphics[width=\columnwidth]{Images/Tonry-1979_Fig3_MEGARA.png}
\caption{Upper panel: Amplitude of the Fourier transform of the MEGARA stellar template (red triangles), this template convolves with a Gaussian function (blue triangles) and the cross correlation of the two previous ones (purple dots). Lower panel: Phase of the same functions mentioned above. The color code is the same as in the upper panel.}
\label{fig:amplitude_phase_MEGARA}
\end{figure}

In this case, we have convolved the stellar template with a Gaussian function with $\sigma$ = 0.18 (MEGARA spectral resolution, HR-I setup) to check if the filter application is correct. Figure \ref{fig:amplitude_phase_MEGARA} shows the phase (upper panel) and the amplitude (lower panel) of the cross-correlation function (the color code is the same as in Fig. \ref{fig:amplitude_phase_MUSE}). Vertical grey lines mark the values used in the band-pass filter (k$_{max}$ = 350 and 2$\cdot$k$_{max}$ = 700 values). This result is similar to that found for MUSE data: at higher wave number values the phase is random and the amplitude of the correlation function is compatible with zero.

\section{Discussion}

\subsection{Method assumptions at different spectral resolutions}\label{sec:suposiciones}
The main purpose of this work is to test the method and assumptions of the classical cross-correlation technique. The principal assumptions of this method are: (i) the broadening function is assumed to be a Gaussian; (ii) the cross-correlation function has a Gaussian behavior and (iii) the amplitude of stellar Fourier transform can also be represented by a Gaussian. 
The first one depends on the particular scientific case to which this methodology is applied since it is based on the idea that the analysed spectrum can be well represented by a star spectrum convolved with a broadening function. This is true, for example, in the case of Seyfert galaxies, circumnuclear star-forming region and HII regions \citep[][]{2006ApJ...641..117G,2007MNRAS.378..163H,1995ApJS...99...67N,2001BSAO...51...11M}.

\begin{figure}
\includegraphics[width=\columnwidth]{Images/Correlation_giants_supergiants_MUSE.png}
\includegraphics[width=\columnwidth]{Images/Correlation_giants_supergiants_MEGARA.png}
\caption{Autocorrelation functions for each star from MUSE (upper panel) and MEGARA (lower panel) data. Giant and supergiant stars are marked with red and blue lines respectively.}
\label{fig:pruebas_corr_stars}
\end{figure}

In order to evaluate the second assumption, we have correlated each stellar template with itself;  this cross correlation is the one carrying the instrumental resolution information and therefore is usually narrower than the star-to-galaxy or galaxy-to-galaxy cross-correlation, hence in very high resolution data it can become too narrow to be adequately represented by a Gaussian function.  

Figure \ref{fig:pruebas_corr_stars} shows the results obtained for all the stars used in this work observed with the MUSE and MEGARA instruments (upper and lower panels respectively). First, we can see that all the stars observed with each of the instruments present similar widths and therefore we are not introducing errors by using their mean value as the star template in our analysis. Then, it is also important to emphasise that MUSE star-to-star cross correlation functions present a Gaussian behaviour and hence the assumption (ii) of the method is correct and therefore applicable at this instrumental resolution. However, the results obtained for MEGARA data show just the opposite: the correlation peak is too narrow and  cannot be approximated by a Gaussian function. Thus, the application of the Tonry \& Davis' method to this kind of data would overestimate the obtained velocity dispersion and any quantity derived from it, including the calculation of dynamical masses.

\begin{figure}
\includegraphics[width=\columnwidth]{Images/amplitude_star.png}
\caption{Amplitude of the Fourier transform of the MUSE (purple line) and MEGARA (green line) stellar templates. Lines made of triangles and dots show the Gaussian fits performed for these two functions respectively.}
\label{fig:pruebas_amp_stars}
\end{figure}

Finally, Figure \ref{fig:pruebas_amp_stars} shows the Fourier transform amplitudes of MUSE and MEGARA templates (in purple and green respectively); triangle and dot lines correspond to the two fitted Gaussian functions. The central peak of each one corresponds to the low frequency variations which are removed from the correlation function after the band-pass filtering. We can see that in both cases a Gaussian amplitude can be assumed, thus the assumption (iii) remains valid at high spectral resolutions.

\subsection{Comparison among methods}

We have verified the spatial resolution range at which the traditional correlation method and the two ones proposed in this work are valid. In order to do that, we have convolved each stellar template spectrum  with known Gaussian functions of different $\sigma$ values, simulating velocity dispersions from 10 km/s to 200 km/s in steps of $\sim$ 10 km/s for MUSE data and from 2 km/s to 100 km/s in steps of $\sim$ 5 km/s for MEGARA data. These spectra have been then cross-correlated with the original template and the output broadening function widths have been calculated. 

\begin{figure}
\includegraphics[width=\columnwidth]{Images/Pruebas_gaussianas_MUSE.png}
\includegraphics[width=\columnwidth]{Images/Pruebas_gaussianas_MEGARA.png}
\caption{Velocity dispersion obtained from the three cross correlation methods studied in this work: the traditional Tonry \& Davis' method (purple dots), the one obtained from [g $\otimes$ g] and [t $\otimes$ t] (red squares) and from [g $\otimes$ g] and [g $\otimes$ t] (green triangles). The results from MUSE and MEGARA data are shown in the upper and lower panel respectively. Error bars have the same length for all the points. }
\label{fig:pruebas_gaussianas}
\end{figure}

The upper panel of Figure \ref{fig:pruebas_gaussianas} shows the comparison between input and output broadening function widths for the three methods proposed in Sec. \ref{sec:teoria} for MUSE data. Purple dots show Tonry \& Davis' method results; red squares show the results using the galaxy-to-galaxy and star-to-star cross correlations, and green triangles show the results in the case of using the galaxy-to-galaxy and galaxy-to-star cross correlations. We can see, at the MUSE spectral resolution the results are entirely compatible within the errors for the tree methods which confirms the results of the previous section that Tonry \& Davis' method can be safely applied at MUSE spectral resolution. Also an asymptotic behavior is found at $\sigma$ = 0.765 \AA\ ($\Delta$v = 26.51 km/s) that probably corresponds to the empirical spectral resolution of the data, that is somewhat lower than the nominal one ($\sigma$ = 2.556 \AA, $\Delta$v = 37.62 km/s). 

The lower panel of the figure shows the same comparison described above but for MEGARA data. In this case, we can see that the two methods which include the star-to-star correlation are not able to recover the input velocity dispersion. At this spectral resolution Tonry \& Davis' method fails being the most discrepant one. In fact it overestimates the derived velocity dispersions by factors ranging from 25 \% at a sigma of 100 km/s to 220 \% at a sigma of 12 km/s. These discrepancies would propagate to any derived quantity. In particular, dynamical masses scale with the square of the velocity dispersion, hence for velocity dispersions of 100 km/s the derived masses would be overestimated by about 50\%; however at low velocity dispersions, as the ones being sought with the use of high resolution data, the derived masses would be overestimated by about one order of magnitude. The best technique to use is the one including the two broader correlation functions, g(n) $\otimes$ g(n) and g(n) $\otimes$ t(n), and, even in this case, discrepancies begin to appear at the lower velocity dispersion values, although they are always within the errors. Therefore this is the recommended method to be applied for high spectral resolution data and results found by the usual application of Tonry \& Davis' method cannot be considered as correct.

\begin{figure}
\includegraphics[width=\columnwidth]{Images/Pruebas_gaussianas_fitt.png}
\caption{Velocity dispersion correction for MEGARA data for the three cross correlation methods studied in this work: the traditional Tonry \& Davis (purple dots) and the ones obtained from [g $\otimes$ g] and [t $\otimes$ t] (red squares) and  [g $\otimes$ g] and [g $\otimes$ t] (green triangles).}
\label{fig:corection}
\end{figure}

We have used Fig. \ref{fig:pruebas_gaussianas} to quantify the differences between measured and real velocity dispersions in the two methods which include the star-to-star correlation. Fig. \ref{fig:corection} shows these differences. A second order polynomial has been fitted for each method, from 10 km/s to 120 km/s, in order to estimate a correction for the MEGARA spectral resolution. For the traditional Tonry \& Davis' method, the correction applied can be calculated using the following equation:
\begin{equation}
\sigma = -1.8339+0.2212\cdot \sigma_{measured} + 0.004858 \cdot (\sigma_{measured})^2
\end{equation}
where $\sigma$ is the real velocity dispersion in km/s and $\sigma_{measured}$ is the measured one in the same units. For the method [g $\otimes$ g] - [t $\otimes$ t] the correction is given by:
\begin{equation}
\sigma = -6.9671+0.6014\cdot \sigma_{measured} + 0.003154 \cdot (\sigma_{measured})^2
\end{equation}

\subsection{Error analysis}

\begin{figure*}
\centering
\includegraphics[width=\textwidth]{Images/Filtered-notfiltered.png}
\includegraphics[width=0.66\textwidth]{Images/masked-notmasked.png}
\caption{upper panels: Cross correlations between frequency filtered and not filtered functions (solid and dotted lines respectively). From left to right are shown: g(v) $\otimes$ t(v), t(v) $\otimes$ t(v), and g(v) $\otimes$ g(v) in units of velocity. Lower panels:  Cross correlations between functions with and without nebular component (dotted and solid lines respectively). Left and right panels show the g(v) $\otimes$ t(v) and  g(v) $\otimes$ g(v) correlation functions.}
\label{fig:pruebas_mask_filt}
\end{figure*}

We have analysed the errors that can occur in two of the steps of the application of the method. The first one is the frequency filtering, and it is shown in the upper panels of Figure \ref{fig:pruebas_mask_filt} that show from left to right the galaxy-to-star, star-to-star and galaxy-to-galaxy cross correlation functions for MEGARA data. Two different lines, solid and dotted, show the results for a filtered spectrum using the values given above (see Sec. \ref{sec:megara_apli}) and a non-filtered one respectively. We can see that the filtering only has an effect on the step level of the correlation peak, hence the velocity dispersion calculations are not affected by errors associated with this part of the procedure.

The second important step is the subtraction of the non-stellar component from the spectrum. If we want to apply this method to galaxies or star clusters the removal of the nebular component is necessary. In order to evaluate the involved errors, we have simulated a nebula ionised by a young star cluster synthesised using the PopStar code \citep{Popstar} with Salpeter's IMF \citep{Salpeter1955} IMF with lower and upper mass limits of 0.85 and 120 M$_\odot$ respectively. We have selected an age of 5 Myr, and constant electron density of 100 cm$^{-3}$. A cluster with 4 $\times$ 10$^{4}$ M$_\odot$ and a nebular and stellar velocity dispersion of 20 km/s has been calculated to represent the simulated clusters. From the H$\beta$ calculated emission line intensity, the equivalent Paschen lines have been obtained with the PyNeb tool \citep{pyneb} for n$_e$(cm$^{-3}$) = 10$^2$ and T(K) = 10$^4$ values using \citet{Storey1995} atomic data.

The lower panels of Figure \ref{fig:pruebas_mask_filt} show the galaxy-to-star and the galaxy-to-galaxy cross correlations respectively, and the correlation without and with nebular emission lines are shown by solid and dotted lines. We can see that the shape of the correlation function peaks and their width are modified, introducing errors in the obtained results. In this particular case, the different measured velocities differ in 4.8 km/s.

\citet{1979AJ.....84.1511T} proposed estimating the internal error of the method from the asymmetric part of the correlation peak, calculating its root mean square. However, after our analysis we can conclude that, at high spectral resolution, this value is not representative of the real measurement error since our main error sources come from the spectrum preparation and the measurement procedure, and not from the quality of the data themselves. Thus, we propose estimating the velocity dispersion errors considering the asymmetries in the correlation peak caused by the entire procedure, calculating the largest and smallest Gaussian width that can be fitted. The corresponding error will be the semi-difference between these two values.

\section{Conclusions}
In this work, we have applied Tonry and Davis' technique  \citep{1979AJ.....84.1511T} to MUSE and MEGARA data in order to evaluate its validity at intermediate and high spectral resolutions. For this purpose, we have used as galaxy spectra convolutions of the stellar template spectrum with known Gaussian functions of different widths. The comparison between the input and output broadening functions shows that the application of Tonry and Davis' method is entirely compatible within the errors for MUSE data but this is not the case for the higher resolution data obtained by MEGARA. This is due the main assumption of the cross-correlation technique not being fulfilled: the cross-correlation function does not present a Gaussian behavior in the star-to-star correlation at this high spectral resolution. 

Therefore, we have adapted Tonry and Davis' method by developing a mathematical equivalent one using the galaxy-to-galaxy correlation instead of the star-to-star correlation which is the one carrying the instrumental resolution information and therefore is the narrower of the three, hence in very high resolution data can become too narrow to be adequately represented by a Gaussian function. At this spectral resolution Tonry \& Davis' method fails entirely, overestimating the derived velocity dispersions by factors up to about 3.5. Taking into account that derived dynamical masses scale with the square of the velocity dispersion, these latter ones would be overestimated by up to a factor of about 10. According to these results the recommended method to be applied for high spectral resolution data is the one using the galaxy-to-galaxy and star-to-galaxy correlations. Results found by the usual application of Tonry \& Davis' method cannot be considered as correct when applied to high spectral resolution data. 

An error analysis shows that the internal errors of the method are unrealistically low and not representative of the real measurement errors. Therefore, we propose estimating errors by calculating the largest and smallest Gaussian widths that can be fitted to the principal correlation peak and using the semi-difference between these two values.

\begin{acknowledgements}
This work has been supported by Spanish grants from the former Ministry of Economy, Industry and Competitiveness through the MINECO-FEDER research grant AYA2016-79724-C4-1-P, the present Ministry of Science and Innovation through research grant PID2019-107408GB-C42 and the National Research Agency through research grant AEI/10.13039/501100011033.

S.Z. acknowledges the support from contract: BES-2017-080509 associated to the first of these grants. 
\end{acknowledgements}

%
\bibliographystyle{aa} 
\bibliography{Article} 
%